\documentclass[twocolumn]{aastex631}
\usepackage{amsmath}
\usepackage{multirow}
\usepackage{amssymb}
\usepackage{graphicx}
\usepackage{color}
\usepackage{float}
\usepackage{comment}

\definecolor{darkgreen}{rgb}{0,0.35,0}
\usepackage{CJK}

\begin{document}
\begin{CJK*}{UTF8}{gbsn}

\title{Radiation Hydrodynamic Simulations of Massive Stars in Gas-rich Environments: Accretion of AGN Stars Suppressed By Thermal Feedback}



\shorttitle{3D RHD Simulations of AGN Stars}
\shortauthors{Chen et al.}

\author[0000-0003-3792-2888]{Yi-Xian Chen (陈逸贤)}
\affiliation{Department of Astrophysical Sciences, Princeton University,  4 Ivy Lane, Princeton, NJ 08544, USA}

\author{Yan-Fei Jiang (姜燕飞)}
\affiliation{Center for Computational Astrophysics, Flatiron Institute, New York, NY 10010, USA}

\author{Jeremy Goodman}

\affiliation{Department of Astrophysical Sciences, Princeton University,  4 Ivy Lane, Princeton, NJ 08544, USA}

\author{Douglas N. C. Lin (林潮)}

\affiliation{Department of Astronomy and Astrophysics, University of California, Santa Cruz, USA}

\affiliation{Institute for Advanced Studies, Tsinghua University, China}


\begin{abstract}
    Massive stars may form in or be captured into AGN disks. 
    Recent 1D studies employing stellar-evolution codes have demonstrated the potential for rapid growth of such stars through accretion up to a few hundred $M_\sun$. 
    We perform 3D radiation hydrodynamic simulations of moderately massive stars' envelopes, 
    in order to determine the rate and critical radius $R_{\rm crit}$ of their accretion process in an isotropic gas-rich environment in the absence of luminosity-driven mass loss. 
    We find that in the ``fast-diffusion" regime where characteristic radiative diffusion speed $c/\tau$ is faster than the gas sound speed $c_s$, the accretion rate is suppressed by feedback from gravitational and radiative advection energy flux, in addition to the stellar luminosity. Alternatively, in the ``slow-diffusion" regime where $c/\tau<c_s$, due to adiabatic accretion, the stellar envelope expands quickly to become hydrostatic and further net accretion occurs on thermal timescales in the absence of self-gravity. 
    When the radiation entropy of the medium is less than that of the star, however, this hydrostatic envelope can become more massive than the star itself. Within this sub-regime, self-gravity of the envelope excites runaway growth. Applying our results to realistic environments, 
    moderately massive stars ($\lesssim 100M_\odot$) embedded in AGN disks typically accrete in the fast-diffusion regime,  
    leading to reduction of steady-state accretion rate 1-2 orders of magnitudes lower than expected by previous 1D calculations and $R_{\rm crit}$ smaller than the disk scale height, except in the opacity window at temperature $T\sim 2000$K. 
    Accretion in slow diffusion regime occurs in regions with very high density $\rho\gtrsim 10^{-9}$g/cm$^3$, and needs to be treated with caution in 1D long-term calculations.
\end{abstract}

\section{Introduction}
\label{sec:intro}

Active Galactic Nuclei (AGN) are bright cosmic beacons \citep{Schmidt1963,Elvis1994} powered by the release of gravitational energy through accretion by supermassive Black Holes (SMBHs) from a surrounding gaseous disk \citep{Lyndenbell1969}.
The structure of these disks is determined by the efficiency of angular momentum transfer \citep{LyndenBell1974,Pringle1981}. 
With the conventional $\alpha$ approach \citep{SS1973}, 
one can only extend a disk heated by MRI turbulence (against radiative cooling) around a $\sim 10^8$ solar mass SMBH to beyond a few light days before Toomre $Q$ parameter \citep{Toomre1964} drops below order unity and sufficient mid-plane density is able to excite gravitational instability (GI) \citep{Paczynski1978,Goodman2003}. 
In outer regions where both GI and MRI turbulence heating is insufficient to balance against radiative cooling in a GI disk, 
intense star formation from disk fragmentation is thought to take place \citep{Levin2003,GoodmanTan2004,Jiang11,Chen2023}. Moreover, these disks and their central SMBHs also commonly coexist with {a dense nuclear cluster of stars} \citep{KormendyHo2013}. 
During their passage through the disks,some of these stars encounter gas drag to sediment onto the disk midplanes \citep{Artymowicz1993,MacLeod2020,WangYH2024}.


These stars deserve special attention in four ways: 
i) A population of Wolf-Rayet and other 
massive main sequence stars exists within 1 pc of Sgr A$^*$ in the Galactic Center, and the alignment of their orbital angular momenta about the black hole suggests origin in a disk 
\citep{Levin+Beloborodov2003,Levin2007,Lu2009} relevant to past AGN activities in the Milky Way. 
The coincidence of these stars’ age $\sim 6 \pm 2$ Myrs \citep{Ghez2003,Paumard2006} and the inferred epoch (a few Myr ago) for
the onset of the Fermi bubble \citep{Su2010} also hint that
these stars may have evolved in the once-active Galactic Center \citep{Zheng2020}; 
ii) When mutual inclination between a massive star and the accretion disk is excited, 
the star might collide and shock-heat the disk during each vertical passage and produce Quasi-periodic Eruption(QPE)-like signals \citep{Linial2023,Tagawa2023};
iii) It has been suggested that merging black holes detected by LIGO-Virgo may find one another in AGN disks \citep{McKernan2012,McKernan2014,Tagawa2020a,Li2021,Samsing2022,Li2022,Epstein-Martin2024} possibly associated with electromagnetic counterparts \citep{graham2020}, 
although merging within star clusters or field binaries remain competing possibilities. 
Stars in AGN disks, 
if numerous, not only serves as progenitors of these compact objects, 
but they themselves will collide with each other and produce bright signatures, 
and might also provide sites for stellar mass BH mergers \citep{Loeb2016,Dai2017}; 
iv) Broad emission lines such as \textsc{Civ, Nv, Oiv} indicate super-solar metallicities in the disks of bright QSOs even at high redshift \citep{Hamann1999,Hamann2002,Nagao2006,Xu2018,Wang+2022,Lai+2022,Huang2023,Floris2024}, 
which may point to in-situ enrichment \citep{Artymowicz1993, Collin+Zahn1999} by star evolution and pollution. 

Regarding implication iii) and iv), 
whether massive stars do provide metallicity to the disk and evolve off the main sequence are dependent on their uncertain evolution in the gas-rich environment of AGN disks. 
Recent works \citep{Cantiello2021,Dittmann2021} applied 1D code MESA \citep{MESA2011,MESA2013,MESA2018,MESA2019} to study stellar evolution in AGN disks over nuclear timescales, 
utilizing simple treatments for stellar accretion and mass loss. 
These studies converge on the idea that under sufficiently high ambient densities, 
stars can attain substantial equilibrium masses (several hundred $M_\odot$) through accretion, 
achieving a balance with mass loss driven by near-Eddington luminosity $\lambda_\star := L_\star/L_{\rm Edd} \approx 1$, both rates being of order $\dot{M} \sim R_\star L_\star/G M_\star$. 
However, 
there is disagreement regarding the the potential for continuous hydrogen supply from the disk to sustain main sequence burning. 
If the stellar radiative zone is quiescent, the inner core will be insulated from disk gas during its nuclear burning and chemical evolution \citep{AliDib2023}. 
However, the diffusion coefficient in the radiative zone of near-Eddington stars is subject to many uncertainties,
and the star may become ``immortal" (continuously burning hydrogen throughout the AGN lifetime) if the hydrogen-rich accreted gas can diffuse efficiently across the radiative zone \citep{Jermyn2022}. Such stars are
unable to generate significant amounts of heavy-element within the AGN lifetime. 


In a series of papers, 
we conduct 3D radiation hydrodynamic simulations of accreting stellar envelopes. 
Our objective is to validate various assumptions utilized in these studies, 
identify their limitations, 
and propose improved modeling prescriptions for 1D evolution of massive stars 
in gas-rich environments over nuclear burning timescales. 
In particular, 
this work focuses on quantifying the accretion rate onto moderately massive stars
embedded in relatively isotropic background environments without significant luminosity-driven mass loss, 
and serves as an initial assessment to determine whether stellar seeds can gain enough mass to reach a few hundred solar mass with $\lambda_\star \approx 1$ within a nuclear burning timescale. 

The accretion prescription currently applied in 1D models \citep{Cantiello2021,Dittmann2021,AliDib2023} is based on the Bondi rate calculated from background density and temperature parameters. 
In addition to this, 
the prescription takes into account radiative feedback from the stellar intrinsic luminosity by multiplying a factor of $1 - \tanh \lambda_\star$ or $(1 -\lambda_\star)^2$. 
However, as we will demonstrate below, this treatment fails to capture 
the extra radiative feedback from energy flux carried by the accretion flow itself, and could significantly overestimate the accretion rate onto AGN stars. 
Moreover, this alteration of accretion rate due to diffusive luminosity 
can only occur in the fast-diffusion limit, which requires the typical radiative diffusion speed $c/\tau$ to exceed the gas sound speed $c_s$, where $\tau$ is optical depth within the Bondi radius. 
When the background environment is dense enough and 
$c/\tau < c_s$, we should enter another regime 
where radiation becomes coupled with gas and can no longer act as a simple reduction to local gravity. 
The resultant impact on accretion across different regimes needs to be investigated by hydrodynamic simulations.

The paper is organized as follows:
In \S \ref{sec:setup}, we introduce our numerical setup.
In \S \ref{sec:results}, 
we present the results obtained from our simulations, deriving approximate scalings for mass accretion rates and critical infall radii as functions of background density and temperature. 
In particular, 
we emphasize the importance of feedback effect from gravitational as well as advective radiation energy flux deposited close to the stellar surface in the fast-diffusion ($c/\tau > c_s$) regime. 
In deriving these scalings, we treat the ambient medium as uniform, isotropic, and non-rotating. 
This is appropriate if $R_{\rm crit}$ is smaller than the vertical scale height of the disk $H$. 
Strong radiative feedback tends to decrease $R_\mathrm{crit}$ and render this assumption self-consistent. 
We also investigate outcomes of nearly adiabatic accretion in the slow-diffusion ($c/\tau < c_s$) regime. 
In \S \ref{sec:implications} 
we discuss the implication of our results for stellar accretion timescales and $R_\mathrm{crit}$ versus $H$ in realistic disk models. 
We also identify disk regions characterized by high accretion rates and/or pronounced anisotropy, 
warranting further investigation. 
Finally, in \S \ref{sec:conclusions}, we summarize our findings and outline future prospects. 
Subsequent works will cover the details of Eddington stellar wind and outflow-accretion interaction in cases where stars may attain sufficient mass through accretion or direct formation, 
as well as explore disk-related anisotropy effects.

\section{3D Numerical Setup}
\label{sec:setup}

\subsection{Equations Solved}

We perform radiation hydrodynamic (RHD) simulations using the standard Godunov method in \texttt{Athena++} \citep{Stone2020}. 
In spherical polar coordinates, we
solve ideal hydrodynamic equations coupled with the time-dependent 
frequency-integrated 
radiation transport equation for specific intensities over discrete angles \citep{Jiang2014,Jiang2021,Goldberg2022,Schultz+2022}:

\begin{equation}
    \frac{\partial \rho}{\partial t}+\boldsymbol{\nabla} \cdot(\rho \mathbf{v})=0
\end{equation}

\begin{equation}
    \frac{\partial(\rho \mathbf{v})}{\partial t}+\boldsymbol{\nabla} \cdot\left(\rho \mathbf{v} \mathbf{v}+\mathbf{P}_{\mathrm{gas}}\right)=-\rho \boldsymbol{\nabla} \Phi
 -\mathbf{G}_{r}
\end{equation}

\begin{equation}
\frac{\partial E}{\partial t}+\boldsymbol{\nabla} \cdot(E+P_{\rm gas}) \mathbf{v}=-\rho \mathbf{v} \cdot \boldsymbol{\nabla} \Phi -cG^0_r
\end{equation}

\begin{equation}
    \frac{\partial I}{\partial t}+c \mathbf{n} \cdot \boldsymbol{\nabla} I=S(I, \mathbf{n})
\end{equation}

In these equations, $\rho$ is the gas density and $\mathbf{v}$ is the 3D flow velocity. 
$\mathbf{P}_{\mathrm{gas}}$ and $P_{\rm gas}$ are the gas pressure in tensorial and scalar form. 
The gas mean molecular weight is $\mu = 0.60 m_p$, consistent with the outer regions of MESA models developed in \citet{Cantiello2021,AliDib2023}, 
with $X = 0.73 , Y= 0.25, Z = 0.02$. 
$E=U_{\rm gas}+\rho v^{2} / 2$ is the sum of gas internal energy $U_{\rm gas}=P_{\rm gas}/(\gamma-1)$ and the kinetic energy $\rho v^2/2$ where $\gamma=5/3$, which is appropriate for $T\gtrsim 10^4$K. 
The source terms $\mathbf{G}_{r}$ and $G^0_r$ are the time-like and space-like components of the radiation four-force \citep{Mihilas1984}. 
$I$ is the frequency-integrated intensity
and $\mathbf{n}$ is the photon propagation direction unit vector. 
Implicit calculations of 
radiation momentum and energy source terms are described in \citet{Jiang2021} and successfully adopted in RHD simulations \citep{Goldberg2022,Schultz+2022,Chen2023}. 
The opacities are interpolated from the OPAL tables \citep{Iglesias1996} and are similar to those applied in MESA for the same value of metallicity.

Unlike purely convective red supergiant stars \citep{Goldberg2022}, 
in our setup of main sequence massive stars the mass in the outer convective envelope is negligible compared to the ``core" mass $M_\star$ enclosed within the inner boundary. 
Therefore, 
gravitational acceleration is taken to be spherically symmetric with $\Phi(r) = -GM_\star/r$ \citep{Schultz+2022}. 

\subsection{Initial Conditions}

We construct 3D isotropic initial conditions from 1D profiles similar to \citet{Jiang2014,Jiang2015,Schultz+2022}.
For massive stars, 
quasi-steady 1D MESA profiles usually predict stellar envelopes to be convective outside an outermost radiative-convective boundary (RCB) at $T\sim 2\times 10^5\mathrm{K}$ where there is a peak in the opacity due to iron.
Simulations reported in this paper are based on a $50\mathrm{M_\odot}$, 1D MESA stellar model with a luminosity of $8.6\times 10^5 L_\odot$ at the end of main-sequence. 
We set the lower boundary of our computational domain in this outermost radiative zone.
To avoid variations in molecular weight and in the abundances used for the opacities, we neglect the compositional gradients that MESA predicts 
for this zone. 
In our initial conditions, we modify the profile of the convective regions so as to support the stellar luminosity purely radiatively; 
this prescription automatically results in convective instability, which our 3D simulations model self-consistently.
The originally radiative zone, however, remains radiative and the velocity fluctuation is significantly lower than in the convective zone.

\subsection{Boundary Conditions}
\label{sec:boundary}

Periodic boundary conditions are implemented in the $\theta$ and $\phi$ direction as described in \citet{Goldberg2022}, 
while the density and temperature are fixed and the velocity is set to zero at the inner boundary in $R$. 
The outer boundary condition in $R$ is free for the velocity, 
while the temperature and density are either free, as for the isolated model, 
or fixed to be constants $T_{\rm out}$ and $\rho_{\rm out}$, respectively. 
The latter boundary values are notionally determined by the disk environment (\S \ref{sec:implications}). 
We set the density and temperature floor to be slightly lower than $\rho_{\rm out}$, 
$T_{\rm out}$ respectively. 
Eventually material inflow from the outer boundary will always ensure $\rho > \rho_{\rm out}$ and $T > T_{\rm out}$ within the simulation domain; the floor values simply help to evolve the accretion system towards a steady state more quickly.
In order to simulate a large scale separation in $r$, 
we simulate from $25 R_\odot$ to  $1.5\times 10^3 R_\odot$ unless stated otherwise, 
while the polar and azimuthal domains span 0.6 rad. 
We add one and two levels of mesh refinement within 300 $R_\odot$ and 150 $R_\odot$ to resolve convection in the inner stellar envelope. 

\subsection{Diagnostics}
\label{sec:diagnostics}

To facilitate analysis of our simulation results, 
we introduce notation for important average variables. 
In a quasi-steady state, 
we define the time and angle-averaged radial profiles as 

\begin{equation}
    \langle X  \rangle (R)= \dfrac{\int X (R, \Omega, t) {\rm d}\Omega {\rm d} t}{\int {\rm d}\Omega {\rm d} t}
    \label{eq:averaging}
\end{equation}
where $\Omega$ is the solid angle. In our analysis, we perform average over the last 100 days of our simulations after they reach steady states. 
We define the density-weighted square of the isothermal sound speed as 

\begin{equation}\label{eq:csdef}
    \langle c_{\rm s}^2 \rangle_\rho = \dfrac{\langle P_{\rm gas} \rangle}{\langle \rho \rangle}
\end{equation}

In adiabatic models where radiative diffusion is weak, 
it's also important to take into account the radiation pressure $P_{\rm rad} = a T^4/3$ where $a$ is the radiation constant, 
and measure the density-weighted total sound speed 

\begin{equation}
    \langle c_{\rm s,tot}^2 \rangle_\rho = \langle c_{\rm s}^2 \rangle_\rho + \langle c_{\rm s, rad}^2 \rangle_\rho
\end{equation}
where the radiation sound speed is 

\begin{equation}
    \langle c_{\rm s, rad}^2 \rangle_\rho = \dfrac{\langle P_{\rm rad}  \rangle}{\langle \rho \rangle}
\label{eq:csrad}
\end{equation}
When the gas and radiation are tightly coupled so as to have a common temperature, the adiabatic sound speed is properly $\sqrt{\Gamma_1(P_\mathrm{rad}+P_\mathrm{gas})/\rho}$, the adiabatic index $\Gamma_1$ varying between $4/3$ and $5/3$ with the ratio $P_\mathrm{rad}/P_\mathrm{gas}$ \citep[e.g.][]{GoodmanTan2004}; but we neglect $\Gamma_1$ here for simplicity.
When determining the critical radius, it's important to compare these profiles with the radial velocity:

\begin{equation}
    \langle v_{R}^2 \rangle_\rho = \dfrac{\langle \rho v_{R}^2  \rangle}{\langle \rho \rangle}
\end{equation}
We also define the radial accretion rate

\begin{equation}
   \langle \dot{M} \rangle = 4\pi R^2 {\langle \rho v_{R}  \rangle}
\end{equation}

The energy flux of the accretion flow is contributed by radiation, gravitational potential, and gas kinetic and thermal energy. 
The radiative luminosity can be further decomposed into a diffusive term $\langle  L_{\rm diff} \rangle= 4\pi R^2 \langle F_{\rm diff} \rangle $ and a term representing the advection of radiative enthalpy $\langle  L_{\rm adv} \rangle= 4\pi R^2 \langle F_{\rm adv} \rangle$. Within our regime of interest, the flow is optically thick enough ($\tau \gg 1$) such that the advective component of the radiative luminosity is $\langle L_\mathrm{adv} \rangle= 4\pi R^2 \langle 4 P_{\rm rad} v_R \rangle$. The ``luminosity" in the form of advected gravitational potential energy is 

\begin{equation}
   \langle  L_{\rm grav} \rangle =  \dfrac{GM_\star}{R} \langle \dot{M} \rangle
\label{eq:lgrav}
\end{equation}
Finally, the luminosity carried by gas thermal and kinetic  energy is 

\begin{equation}
   \langle  L_{\rm gas} \rangle = 4\pi R^2  \langle (U_{\rm gas} + \rho v^2 ) v_R \rangle 
\end{equation}
This last term usually plays a minor role in the energy budget, being at most comparable to the gravitational potential term.


    

We define the sum of scattering and absorption opacity as $\kappa = \kappa_{\rm s} + \kappa_{\rm a}$.
The dimensionless measure of diffusive luminosity, characterising the effect of reduced gravity, is measured as

\begin{equation}
    \lambda_{\rm diff} = \left\langle \dfrac{L_{\rm diff}}{L_{\rm Edd}} \right\rangle =  \dfrac{R^2}{GM_\star c}  \langle {F_{\rm diff}}{\kappa} \rangle
\end{equation}
Here we've introduced the local Eddington luminosity $L_{\rm Edd} = 4\pi GM c/\kappa $
and Eddington ratio $\lambda_{\rm diff}$ based on the local opacity $\kappa(r)$ rather than the (constant) electron-scattering opacity.
Finally, we define the flux or luminosity-weighted opacity

\begin{equation}
    \langle \kappa \rangle_{L} = \dfrac{\langle {F_{\rm diff}}{\kappa} \rangle}{\langle {F_{\rm diff}}\rangle}
\end{equation}
such that $\lambda_{\rm diff}\propto \langle \kappa \rangle_{L} {\langle {F_{\rm diff}}\rangle}$.

\section{Results}
\label{sec:results}


\begin{table*}
\centering
\begin{tabular}{ccccc|cc}
\hline
Model&$\rho_{\rm out}$(cgs) & $T_{\rm out}$ (K) &$t_{\rm final}$ (days) & predicted $\dot{M} (M_\odot$/yr) & simulated $\dot{M} (M_\odot$/yr)  & Comment \\\hline
\texttt{Iso}& -- & -- & 100 &  -- & -- & Isolated Model 
\\\hline 
\texttt{D1e-12T6e4}& $10^{-12}$ & $6\times 10^4$ & 300 & 0.00024 & 0.00049  & enthalpy feedback, Figure \ref{fig:comparison_profiles}
\\\hline
\texttt{D1e-11T6e4}& $10^{-11}$ & $6\times 10^4$ & 300 & 0.0023  & 0.0037  &enthalpy feedback, Figures \ref{fig:snapshot}, \ref{fig:spacetime}, \ref{fig:fiducial_1Dprofiles}, \ref{fig:fiducial_1Denergyprofiles}
\\\hline
\texttt{D1e-10T6e4}& $10^{-10}$ & $6\times 10^4$ & 300 &  0.014  & 0.024  &combined feedback, Figure \ref{fig:comparison_profiles}
\\\hline
\texttt{D1e-10T3e4}& $10^{-10}$ & $3 \times 10^4$ & 300 &  0.028  & 0.036 &gravitational feedback, Figure \ref{fig:comparison_profiles},\ref{fig:effect_of_kappa}
\\\hline
\texttt{D1e-10T6e3}& $10^{-10}$ & $6 \times 10^3$ & 300 &  1.7 & 1.2  & minimal feedback, Figures \ref{fig:effect_of_kappa}, \ref{fig:lowkappa_1Dprofiles}
\\\hline
\texttt{D1e-9T6e4}& $10^{-9}$ & $6 \times 10^4$ & 300 &  --  & -- & slow diffusion, Figures \ref{fig:spacetime_adi}, \ref{fig:adi_1Dprofiles}
\\\hline
\texttt{D1e-9T6e4\_adi}& $10^{-9}$ & $6 \times 10^4$ & 3000 &  --  & -- & adiabatic run, Figures \ref{fig:adi_1Dprofiles}, \ref{fig:spacetime_adi_woRT},\ref{fig:longterm_energy}
\\\hline
\end{tabular}
\caption{List of model names and parameters used in our simulations. 
Apart from the isolated case, $\rho_{\rm out}$ and $T_{\rm out}$ indicate the density and temperature fixed at the outer boundary. 
The predicted accretion rate is derived from Equation \ref{eqn:quartic_for_Mdot} assuming fast-diffusion limit and compared to measured accretion rates measured from simulations that reached a quasi-steady state. 
For slow-diffusion runs (the last two rows) no steady accretion rate is measured.}
\label{tab:parameters}
\end{table*}


In run \texttt{Iso} we evolve a stellar atmosphere in isolation (with a zero-gradient outer boundary in density and pressure, 
and an outflow boundary condition in velocity) that reaches a quasi-steady state at $t=100$ days.
Run \texttt{Iso} evolved into a steady state with a turbulent convective envelope extending up to $\sim 75R_\odot$ and a quiescent radiative zone below $\sim 30R_\odot$, 
roughly consistent with the MESA profile. We did not observe any persistent outflow driven by the radiative force at the iron opacity bump,
such as may drive the winds of in Wolf-Rayet stars \citep{Grassitelli2018}. 
Henceforth, we focus solely on the impact of accretion on this moderate-mass stellar model. 
Environmental parameters for other accreting runs (with fiducial boundary 
condition for various values of $T_{\rm out}$ and $\rho_{\rm out}$) are listed in Table \ref{tab:parameters}.
Most of our simulations are run to a quasi-steady state at $t=300$ days.

\begin{figure}
    \centering
    \includegraphics[width=0.52\textwidth]{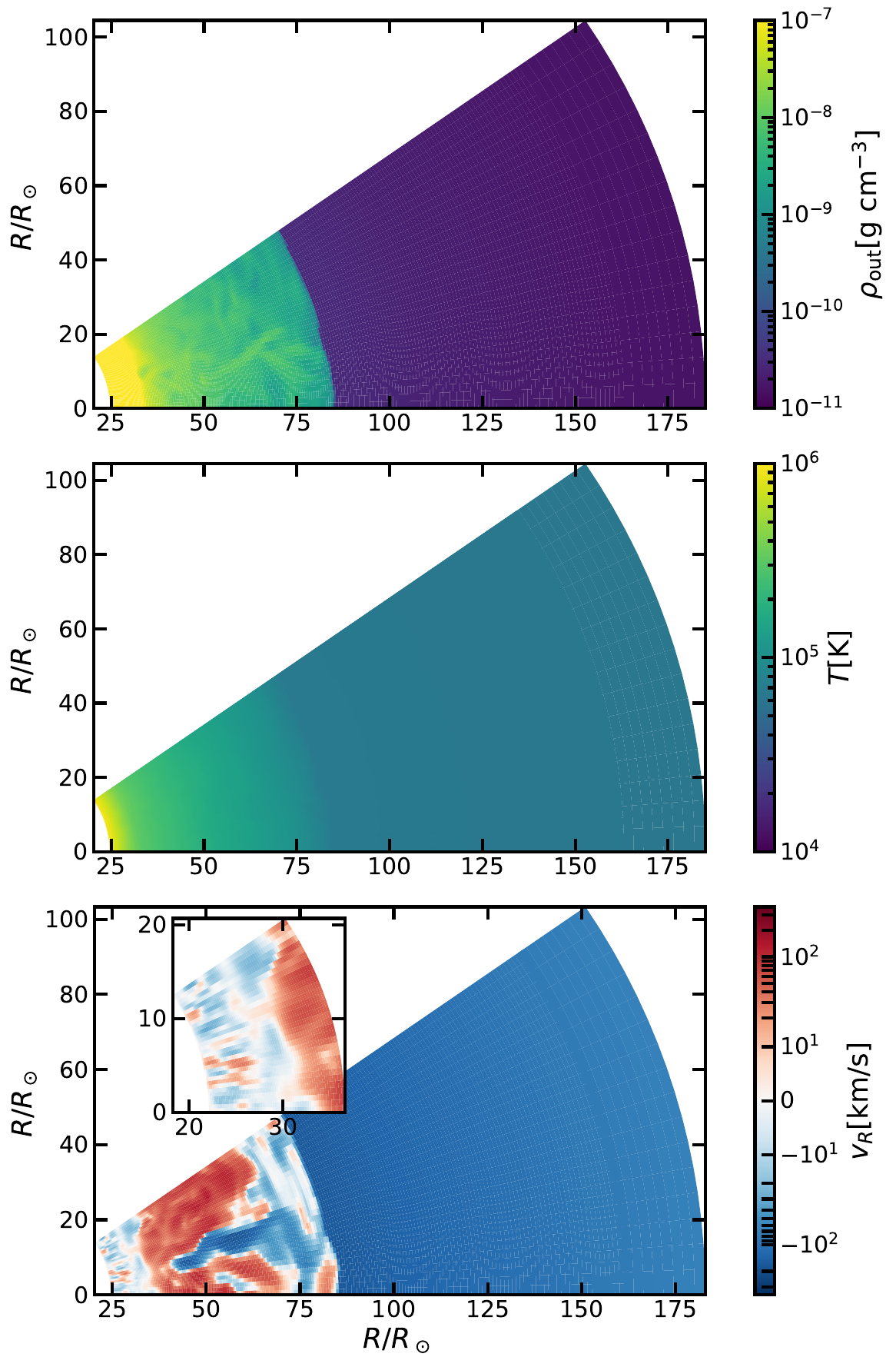}
    \caption{A typical snapshot of density, temperature and radial velocity distributions in the midplane for our fiducial run \texttt{D1e-11T6e4} in quasi-steady state. Note that only the $R < 150R_\odot$ ($40 
    R_\odot$) regions are shown (in the insert) to emphasize the inner envelope.
    }
\label{fig:snapshot}
\end{figure}

\begin{figure*}
    \centering
    \includegraphics[width=0.75\textwidth]{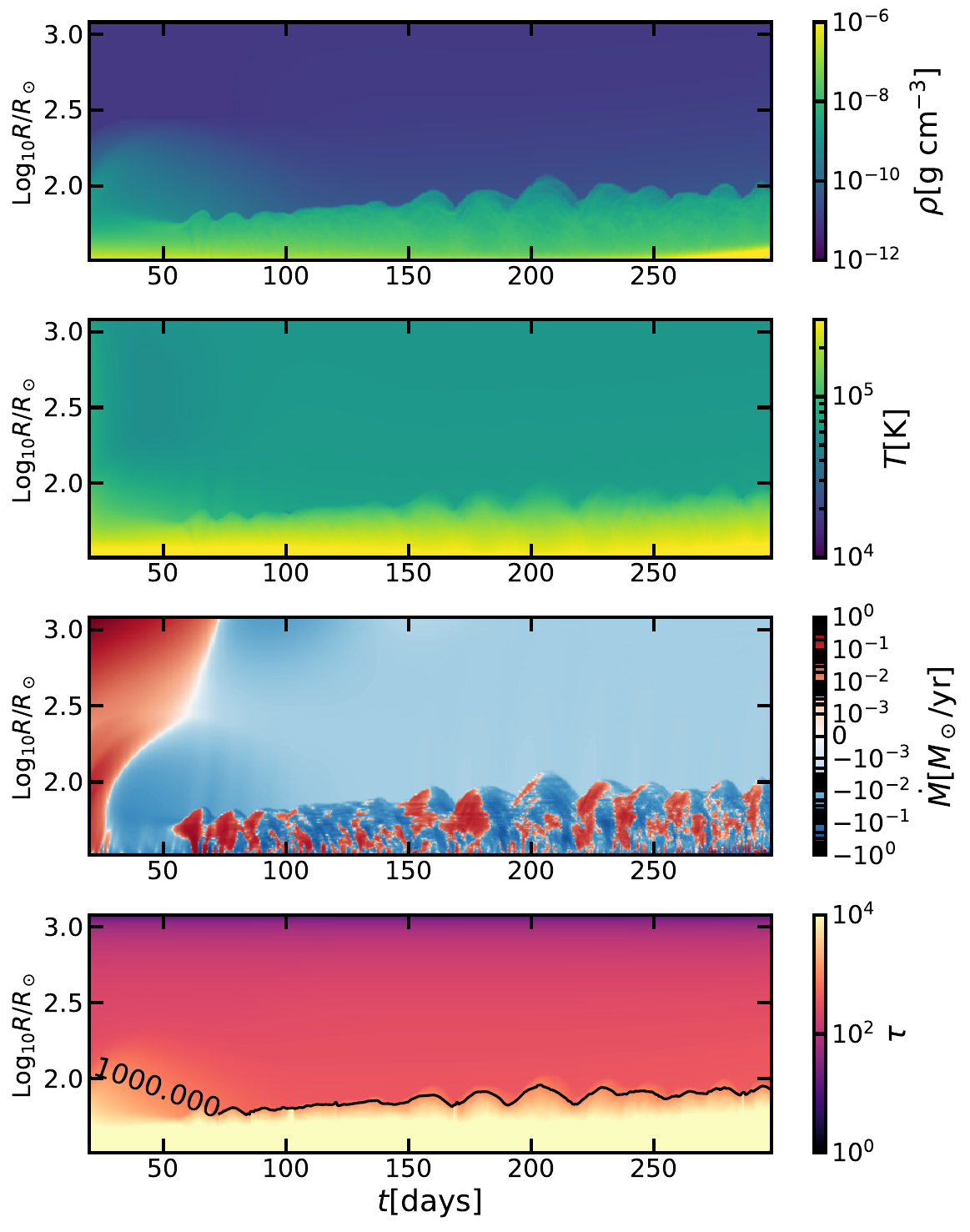}
    \caption{Angle-averaged radial distributions of density, temperature, {radial accretion rate} and optical depth as functions of time (horizontal axis) for fiducial run \texttt{D1e-11T6e4} over 50 days. 
    The $\tau=1000$ surface is indicated as a black contour in the bottom panel that roughly follows the boundary of the convective envelope.}
\label{fig:spacetime}
\end{figure*}

\subsection{Fiducial Run}

Figure \ref{fig:snapshot} shows a typical snapshot of midplane $(\phi = 0)$ density, 
temperature and accretion rate distributions 
in the fiducial accreting run \texttt{D1e-11T6e4}, in which $\rho_{\rm out} = 10^{-11}{\rm g/cm}^3$ and $T_{\rm out} = 6\times 10^{4}$K (or $c_{\rm s, out} = 29$km/s), 
after it has reached a quasi-steady state. 
The convective envelope extends outward to approximately $70-80 R_\odot$. 
At larger radii, 
there is relatively isotropic mass inflow. 
The density and temperature profiles of the convection zone closely resemble those of the starevolved in isolation.
The outer edge of the convection zone
acts as a kind of boundary layer where the accretion flow stalls and its mass is redistributed into the envelope as a whole. 
The primary contrast with the isolated model arises in the regions exterior to the convective envelope, 
where a supersonic accretion flow is established. 
Figure \ref{fig:spacetime} shows the time evolution of the averaged radial profiles of density, temperature, 
accretion rate as well as the optical depth integrated from the outer boundary, 
with the vertical axis being logarithmic. 
This shows that {by the end of the simulation}, 
the stellar profiles indeed
attain a quasi-steady state. 

The bottom panel of Figure \ref{fig:spacetime} shows that the optical depth from the outer boundary to the convective layer at $R_*$ is $\sim 10^3$, much smaller than $c/c_{\rm s, out}\sim 10^5$, which means that characteristic photon diffusion timescale is much shorter than the sound crossing timescale. 
In this regime, diffusive luminosity acts as reduction of gravity, and the radiative energy density is minimally advected with the matter. 
Averaging over the quasi-steady phase, 
we can study 
the radial profiles and accretion rate of this stellar accretion solution.

\begin{figure}
    \centering
    \includegraphics[width=0.48\textwidth]{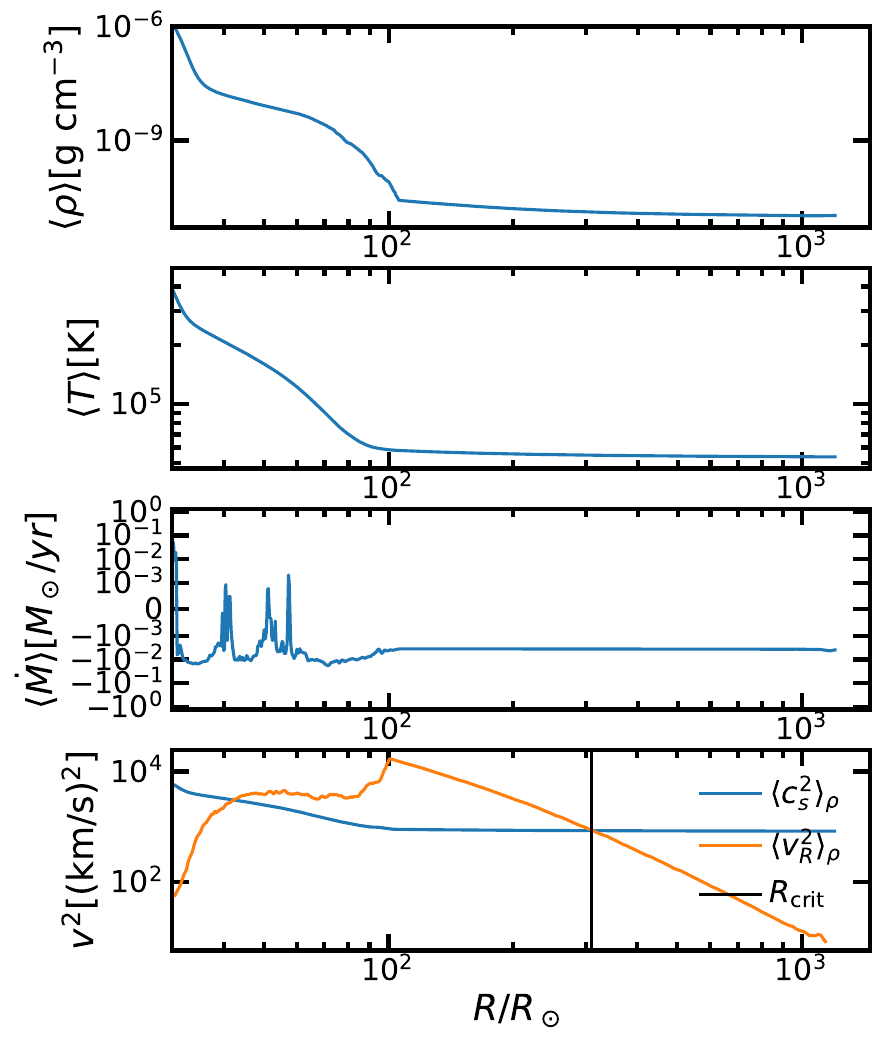}
    \caption{From top to bottom:  $\langle \rho \rangle, \langle T \rangle, \langle \dot{M} \rangle$ as well as $\langle c_{\rm s}^2 \rangle_\rho$ (blue) v.s. $\langle v_{r}^2\rangle_\rho$ (red) and $R_{\rm crit}$ (black) radial profiles from the fiducial run, averaged over the last 100 days of the simulation. The vertical black line indicates where infall velocity becomes supersonic with respect to the sound speed based on the gas pressure ($c_\mathrm{s}$). The details of averaging see \S \ref{sec:diagnostics}. 
    }
\label{fig:fiducial_1Dprofiles}
\end{figure}

\begin{figure}
    \centering
    \includegraphics[width=0.42\textwidth]{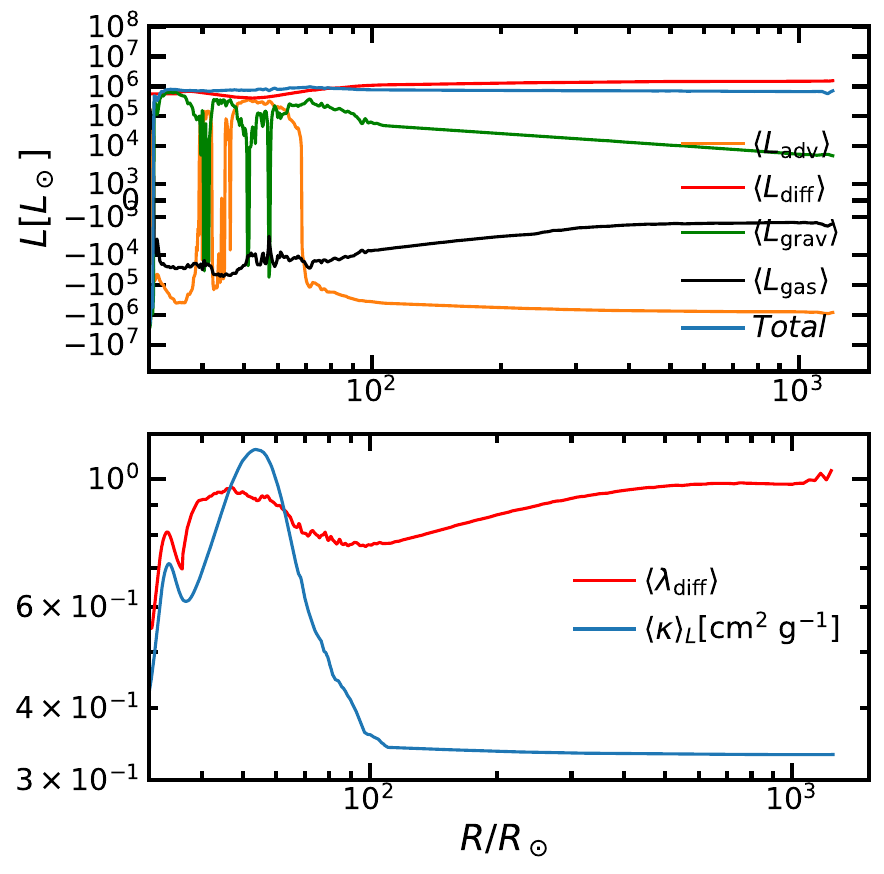}
    \caption{Top panel: average radial profiles for different energy fluxes in the fiducial run. Lower panel: average radial profiles for the local Eddington factor $\langle \lambda_{\rm diff} \rangle$ (red) as well as diffusive flux-weighted opacity $\langle \kappa \rangle_{\rm L}$ (blue), 
    which highlights that contribution of enhanced $\langle \lambda_{\rm diff} \rangle$ at large radius
    is due to the effect of energy feedback (increase in $\langle L_{\rm diff} \rangle$) instead of the opacity bump  
    (nearly constant $\langle \kappa \rangle_{\rm L}$).
    For definition of these variables and details of averaging see \S \ref{sec:diagnostics}.}
\label{fig:fiducial_1Denergyprofiles}
\end{figure}

Figure \ref{fig:fiducial_1Dprofiles} shows the angle-averaged $\langle \rho \rangle, \langle T \rangle, \langle \dot{M} \rangle$ 
as well as $\langle c_{\rm s}^2 \rangle_\rho$ v.s. $\langle v_{r}^2\rangle_\rho$. 
These quantities are also averaged over the time interval between
200-300 days.
We highlight the critical radius 
(black vertical line) where infall velocity becomes supersonic with respect to the local gas sound speed (Equation \ref{eq:csdef}).
At the densities and optical depths of this fiducial run, radiative diffusion carries most of the total luminosity outside the convection zone (Figure \ref{fig:fiducial_1Denergyprofiles}) with a relatively shallow temperature gradient (Figure \ref{fig:fiducial_1Dprofiles}).  
Meanwhile, 
the density also only slightly rises from $\rho_{\rm out}$ within the critical radius 
and abruptly rises close to the boundary of the convective envelope. 
The steady mass accretion rate up to the convective envelope $\dot{M} \approx 3\times 10^{-3} M_\odot$/yr 
is consistent with a modified trans-sonic radius of $R_{\rm crit}$ where the infall velocity crosses the sound speed, 
or $\dot{M} \sim 4\pi R_{\rm crit}^2 \rho_{\rm out} c_{\rm s, out}$ 
since the density and temperature at $R_{\rm crit}$ is close to the outer boundary value. 


It has been hypothesized in 1D studies \citep{Dittmann2021,AliDib2023} that a stellar (diffusive) luminosity of $L_\star = L_{\rm diff}= \lambda_\star L_{\rm Edd}$ introduces a reduction in gravity, 
causing the {Bondi radius} to scale down as

\begin{equation}
    R_{\rm B} = (1-\lambda_\star) \dfrac{GM_\star}{2c_{\rm s, out}^2} 
\end{equation}
where $GM_\star/2c_s^2$ is the classical Bondi radius when the gas is nearly isothermal or efficiently cooled by radiation. One may also approximate the reduction factor $1-\lambda_\star$ as some function of stellar mass \citep{WangJM2023}.
Neglecting order-unity coefficients involving the effective adiabatic index, the modified accretion rate would be
\begin{equation}
    \dot{M}_{\rm B} \approx 4\pi  R_{\rm B}^2 \rho_{\rm out} c_{\rm s, out}
    \label{eqn:classical}
\end{equation}


For our stellar model with $\lambda_\star = 0.6$ and taking the fiducial boundary parameter, we estimate $R_{\rm B} \approx 2074 R_\star$
and $\dot{M}_{\rm B} \approx 0.11 M_\odot$/yr. 
However, 
our measured values are significantly smaller.
This is because the above simple analysis assumes a constant Eddington ratio $\lambda_{\rm diff} = \lambda_\star$ from the stellar surface to the critical radius. 
In reality, 
both $L_{\rm diff}$ and $L_{\rm Edd}$ vary with radius due to other energy flux components in the accretion flow that provide strong feedback. 
To illustrate our point, 
in Figure \ref{fig:fiducial_1Denergyprofiles} we plot the energy budget 
as well as Eddington ratio measured from the simulations. 

Across the simulation domain, there are significant variations in $\lambda_{\rm diff}$ due to a number of factors.
First, 
the trend in $\langle \kappa \rangle_{L}$ is reasonably clear such that we can separate the effects of convection from the feedback due to other energy fluxes. 
Within the convective layer,  
$\langle \kappa \rangle_{L}$ peaks at the iron opacity bump, 
and $L_{\rm diff}$ must decrease in order to remain below the local Eddington luminosity, 
maintaining local $\lambda_{\rm diff}$ that is close to but no more than unity. 
To maintain radially constant total luminosity (as required for a quasi-steady state), 
the advective energy fluxes in gas and/or radiation must rise to make up for the reduction in $L_{\rm diff}$, accompanied by a net energy-weighted velocity outwards, 
as in standard convection theory \citep{Jiang2015}.

The maximum value of $\lambda_{\rm diff}$ in the convective zone is sensitive to the details of turbulence in the convective layer and is hard to predict. 
However, 
this deviation of $\lambda_{\rm diff}$ from $\lambda_{\star}$ is \textit{not} what determines the accretion rate at $R_{\rm crit}$. 
This is because, beyond an effective outer boundary of the convective zone (which we term $R_\star$, and measured to be $\sim 75 R_\odot$ in our simulations), 
$\langle \kappa \rangle_{L}$ again relaxes to the constant electron-scattering opacity $\kappa_{\rm e}$
whereas it is within the range $R_\star < R < R_{\rm crit}$ that the steady accretion flow is established. 
From Figure \ref{fig:fiducial_1Denergyprofiles} we see that as convection dies out beyond $R_\star$, the diffusive luminosity $L_{\rm diff}$ converges to a value close to its value at the inner boundary, $L_\star$. 
This is also a general feature for non-accreting massive stars with outer convective envelopes \citep{Jiang2015}.

Decoupling the effect of varying $\langle \kappa \rangle_{L}$ in the convective layer, 
let us take a closer look at the $L_{\rm diff}$ and $\lambda_{\rm diff}$ profiles \textit{outside} $R_\star$.  
In this region, 
diffusive luminosity rises from $L_\star$ to keep total luminosity (blue line) constant as other components of the luminosity decreases due to the existence of $\dot{M}$. 
In particular, 
the positive $\langle L_{\rm grav}\rangle $ decreases as $\propto 1/r$ for a constant accretion rate (green line), 
while the negative $\langle L_{\rm adv}\rangle $ decreases as $\propto -1/\rho$ for a nearly isothermal profile (yellow line). 
The contribution from $\langle L_{\rm gas} \rangle$ (black line) is minor compared with those two. 
The physical interpretation is that gravitational potential energy and radiation enthalpy are deposited along the accretion flow and carried out by extra diffusive radiation luminosity, 
the former being analogous to the Eddington-accretion feedback mechanism for black holes. 
Since $L_{\rm Edd}$ is nearly constant, 
$\lambda_{\rm diff}$ increases solely due to $L_{\rm diff}$ 
until it reaches some value much closer to unity than $\lambda_\star$ at a radius of $R_{\rm crit}$, 
leading to a much stronger feedback than predicted by Equation \ref{eqn:classical}.

\subsection{Quantifying Feedback}
\label{sec:feedback}

In our first attempt to quantify feedback, 
let us assume the opacity from the surface of convective zone towards the critical radius is nearly constant at the electron-scattering value $\kappa_{\rm e}$, 
as in our fiducial case. 
Under this approximation, 
$L_{\rm Edd}$ from $R_\star$ to $R_{\rm crit}$ is also a constant. A constant total luminosity in the optically thick region requires 

\begin{equation}
\begin{aligned}
    &L_{\rm diff} (R_\star) + L_{\rm adv} (R_\star) + L_{\rm grav} (R_\star) + L_{\rm gas} (R_\star) \\=& L_{\rm diff} (R_{\rm crit}) + L_{\rm adv} (R_{\rm crit}) + L_{\rm grav} (R_{\rm crit}) + L_{\rm gas} (R_{\rm crit})
    \end{aligned}
\end{equation}

This equation can be approximately written as

\begin{equation}
    L_{\rm diff} (R_{\rm crit}) = L_{\rm diff} (R_\star) + \left(f_1 \dfrac{GM_\star}{R_\star} + f_2 \dfrac{4 a T_{\rm out}^4 }{\rho_{\rm out}}\right) \dot{M}
\end{equation}
The term involving $f_1$ represents
$L_{\rm grav}(R_{\rm crit})$ scaled by $L_{\rm grav}(R_{\star})$, on the assumption that $\dot M$ is constant with radius.
The term in $f_2$ scales the (negative) luminosity associated with 
the  inward advection of 
radiative enthalpy by its value at the outer boundary. Since a supersonic inflow extends from $R_{\rm crit}$ to the stellar surface at $R_*$, the gas-enthalpy term is small compared to the gravitational-potential terms. 


The overall reduction parameter for gravity becomes

\begin{equation}
    \lambda_{\rm crit} := \lambda_{\rm diff} (R_{\rm crit}) = \lambda_\star + \dfrac{\dot{M} }{L_{\rm Edd}}\left( f_1\dfrac{GM_\star}{R_\star} + f_2 \dfrac{aT_{\rm out}^4}{\rho_{\rm out}}\right) 
\end{equation}
in steady-state, the 1D momentum and continuity equations can be combined to obtain

\begin{equation}
    \left(\dfrac{c_s^2 - v_R^2}{\rho} \right) \dfrac{d\rho }{d R} = \dfrac{2v_R^2 }{R} - (1-\lambda_{\rm diff}) \dfrac{GM}{R^2}-\dfrac{d c_\mathrm{s}^2}{d R}
    \label{eqn:momentum}
\end{equation}
such that when both ${d c_\mathrm{s}^2}/{d R} \ll { c_\mathrm{s}^2}/{R}$ and gas is cooled by radiation relatively efficiently, the critical radius and $\dot{M}$ are determined self consistently by requiring:

\begin{equation}
   R_{\rm crit} =  (1-\lambda_{\rm crit}) \dfrac{GM_\star}{2c_s^2}
\end{equation}

\begin{equation}
    \dfrac{\dot{M}}{\dot{M}_{\rm B}} = \left(\dfrac{R_{\rm crit}}{R_{\rm B}}\right)^2.
\end{equation}
This expression translates to a quadratic equation for $\dot{M}$:

\begin{equation}
    \dfrac{\dot{M}}{\dot{M}_{\rm B}} = \left(1 - \dfrac{\dot{M}}{\dot{M}_{\rm fb}}\right)^2
    \label{eqn:quartic_for_Mdot}
\end{equation}
where we define the ``feedback" accretion rate to be

\begin{equation}
    \dot{M}_{\rm fb}^{-1} = f_1\dot{M}_{\rm g}^{-1} + f_2 \dot{M}_{\rm r}^{-1}
    \ \ \ \ \ \ {\rm with}
\end{equation}

\begin{equation}
    \dot{M}_{\rm r} = (1-\lambda_\star) \dfrac{L_{\rm Edd} \rho_{\rm out}}{4 aT_{\rm out}^4} \ \ \ \ \ \ {\rm and}
    \label{eqn:Mdotenthalpylimit}
\end{equation}

\begin{equation}
    \dot{M}_{\rm g} =  (1-\lambda_\star)\dfrac{L_{\rm Edd} R_\star}{GM_\star} = (1-\lambda_\star) \dfrac{4\pi R_\star c}{\kappa}.
    \label{eqn:Mdotgravitylimit}
\end{equation}
With Equations (\ref{eqn:classical}) and (\ref{eqn:quartic_for_Mdot}),
$\dot{M}$ and $\dot{M}_{\rm B}$ can be determined as functions of 
$\rho_{\rm out}, T_{\rm out}$. Effectively, 
the solution to Equation \ref{eqn:quartic_for_Mdot} 
states that the mass accretion rate is capped by feedback limits from both gravitation and radiation enthalpy flow such that

\begin{equation}
    \dot{M} \approx \min[\dot{M}_{\rm B}, \dot{M}_{\rm fb}] \approx \min[\dot{M}_{\rm B}, \dot{M}_{\rm g}/f_1, \dot{M}_{\rm r}/f_2]
\label{eq:dotMmin}
\end{equation}

\begin{figure}
    \centering
    \includegraphics[width=0.48\textwidth]{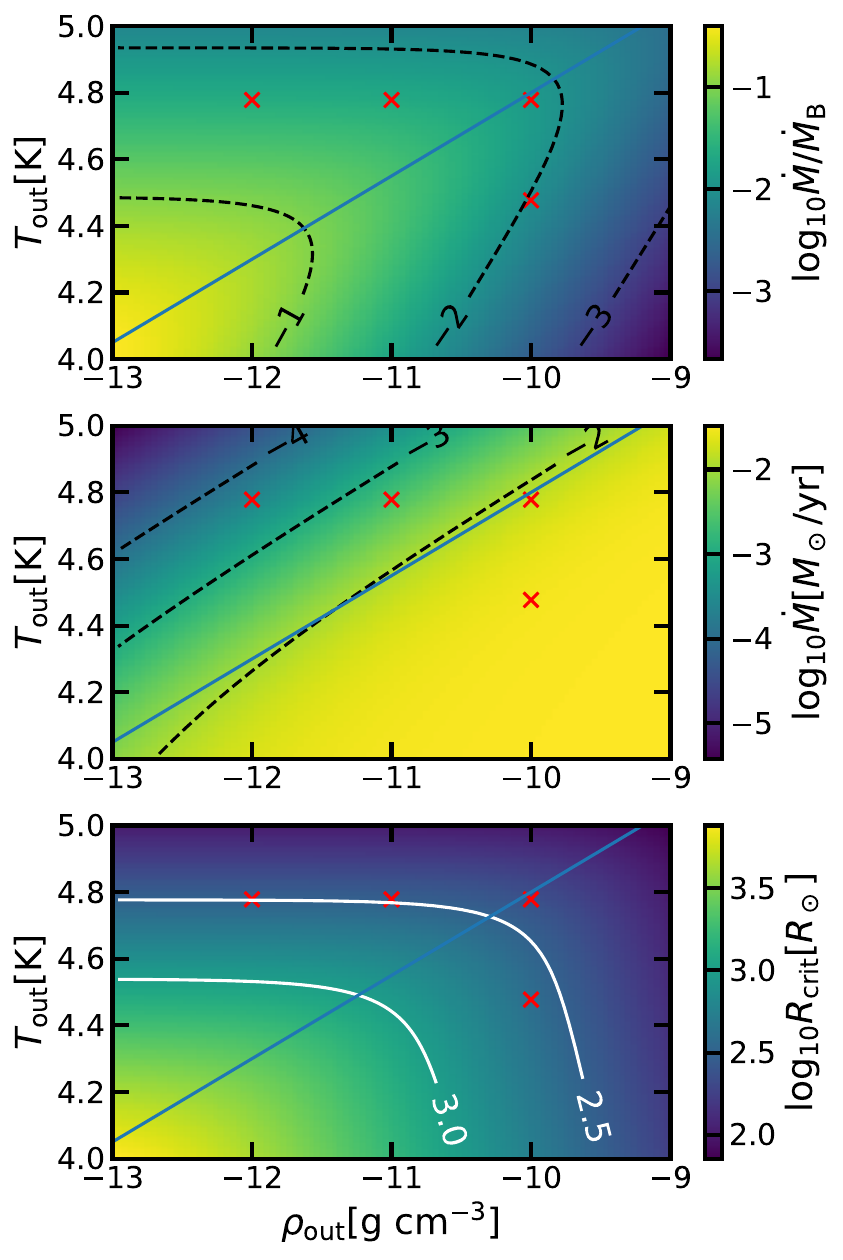}
    \caption{Solutions to Equation \ref{eqn:quartic_for_Mdot} as functions of $T_{\rm out}$ and $\rho_{\rm out}$. The color map and black dashed contour lines in top and middle panels present $\dot{M}$ in terms of Bondi accretion rate and solar mass per year, respectively.  The color map and solid white contour lines in the lower panel present $R_{\rm crit}$ in terms of solar radius. The separatrix between enthalpy feedback and gravitational feedback $c_{\rm s, rad, out} \sim v_{\rm esc}$ is highlighted by blue solid line. The red crosses indicate the simulations listed in Table~\ref{tab:parameters}.}
\label{fig:simple_parameter_space}
\end{figure}

In order to show the transition between enthalpy-limited and gravity-limited accretion, 
we plot the solution of 
Equation \ref{eqn:quartic_for_Mdot} over the $(\rho_{\rm out}, T_{\rm out})$ parameter space in Figure \ref{fig:simple_parameter_space}, 
assuming $f_1 = f_2 = 1, R_\star = 75R_\odot$. 
First, we observe that for this range of $(\rho_{\rm out}, T_{\rm out})$ the accretion rate is always reduced to $\ll \dot{M}_{\rm B}$, 
by one feedback or another. 
Second, the two feedback regimes are clearly distinguished by the separatrix (plotted with blue line)

\begin{equation}
    \dfrac{4 a T_{\rm out}^4}{\rho_{\rm out}} \sim \dfrac{GM_\star}{R_\star} \quad \text{or} \quad  c_{\rm s, rad, out} \sim  v_{\rm esc} 
\end{equation}

where $v_{\rm esc} $ is the escape velocity at $R_\star$. When $c_{\rm s, rad, out}$ or the radiation sound speed at the outer boundary is larger, 
the enthalpy feedback is dominant, 
and $\dot{M} \sim \dot{M}_{\rm r} $ is inversely proportional to the radiation pressure, 
while the ratio to Bondi accretion is a function of temperature only (this is because both $\dot{M}_{\rm B}$ and $\dot{M}_{\rm r}$ are proportional to density). 
Since $R_B$ is independent of density, 
the absolute radius of $R_{\rm crit}$ is also only a function of temperature. Note that this analysis assumes that even when radiative diffusion is fast, we are still in the optically thick regime ($c/c_s > \tau \gg 1$) such that radiation-enthalpy advection is important, which is the case with our fiducial simulation (and the suite of simulations we describe in \S\ref{sec:verify_scaling}).

When $v_{\rm esc} $ becomes larger than the radiation sound speed, gravitational feedback becomes dominant and the accretion rate relaxes to $\dot{M}_{\rm g}$ [\eqref{eqn:Mdotgravitylimit}].
The ratio $\dot{M}/\dot{M}_{\rm B} $ simply scales as the inverse of Bondi accretion rate, and $R_{\rm crit} \propto \dot{M}_{\rm B}^{-1/2} R_B$.


\subsection{Varying Accretion Parameters}
\label{sec:verify_scaling}

To verify the scaling relations drawn from the above discussion, 
we present results from runs \texttt{D1e-12T6e4, D1e-10T6e4, D1e-10T3e4}, 
whose parameters are plotted in Figure \ref{fig:simple_parameter_space} as red crosses, 
on top of the fiducial \texttt{D1e-11T6e4} parameter. 

\begin{figure}
    \centering
    \includegraphics[width=0.45\textwidth]{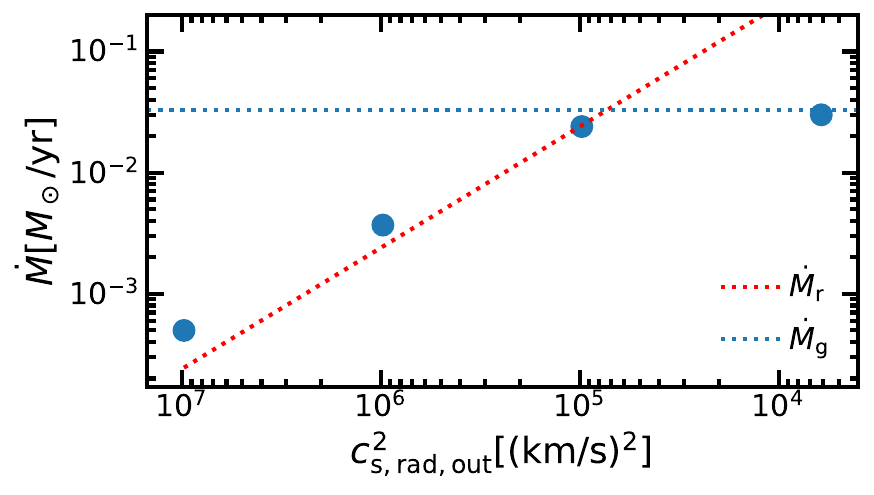}
    \caption{Data points from left to right: 
    measured stellar accretion rates from Models \texttt{D1e-12T6e4}, \texttt{D1e-11T6e4}(fiducial), \texttt{D1e-10T6e4}, 
    \texttt{D1e-10T3e4}. 
    The horizontal axis shows the square radiation sound speed defined by the environment parameters. 
    The red and blue dotted lines show upper limits of the accretion rate assuming an electron scattering opacity, 
    $R_\star = 75R_\odot$, 
    and $f_1 = f_2 = 1$ in Equations \ref{eqn:Mdotenthalpylimit} 
    and \ref{eqn:Mdotgravitylimit}.}
\label{fig:data_points}
\end{figure}

As we progress to larger $\rho_{\rm out}$ and lower $T_{\rm out}$, 
the most direct comparison is to plot the accretion rate measured in physical units 
as a function of $c_{\rm s, rad, out }^2$  
($= aT_{\rm out}^4/3 \rho_{\rm out}$). 
From Figure \ref{fig:data_points} we plot the simulation results as data points against scalings from Equations \ref{eqn:Mdotenthalpylimit} and \ref{eqn:Mdotgravitylimit}, 
assuming the Eddington luminosity is determined by an electron scattering opacity $\kappa_{\rm e}$ at the critical radius and taking $f_1 = f_2 = 1$. 
We observe that accretion rate first increases as $\dot{M}_{\rm r}$ and then relaxes to a constant $\dot{M}_{\rm g}$, 
consistent with the scalings (Equation \ref{eq:dotMmin}) proposed in \S\ref{sec:feedback}.
Physically, Figure \ref{fig:data_points} 
illustrates a transition from the enthalpy feedback limit to the gravitational feedback limit, 
as the magnitude of variation in $\langle L_{\rm adv} \rangle $ diminishes relative to that in $\langle L_{\rm grav} \rangle $.

\begin{figure}
    \centering
    \includegraphics[width=0.48\textwidth]{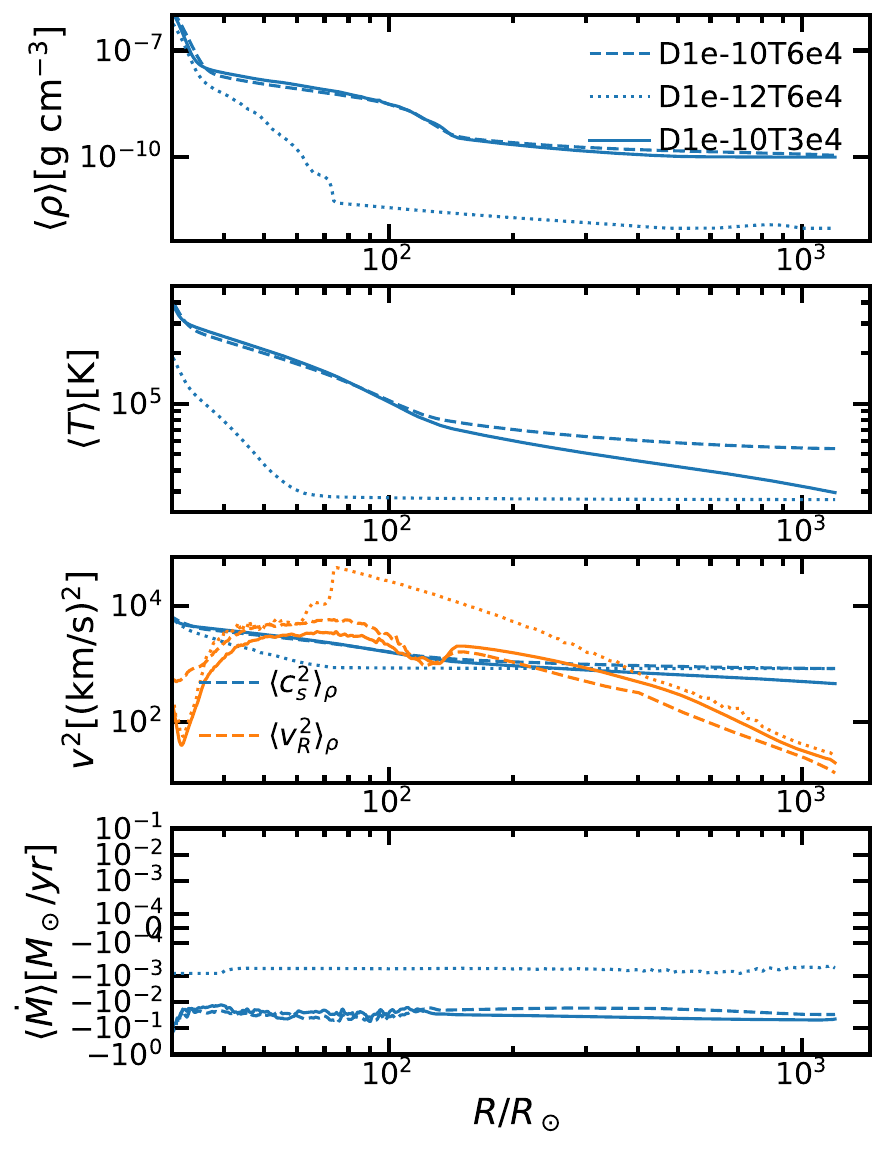}
    \caption{Similar to Figure \ref{fig:fiducial_1Dprofiles} but for runs \texttt{D1e-12T6e4, D1e-10T6e4, D1e-10T3e4} (dotted, dashed, solid lines).}
\label{fig:comparison_profiles}
\end{figure}

We plot the 1D average profiles for runs \texttt{D1e-12T6e4, D1e-10T6e4, D1e-10T3e4} similar 
to the fiducial case in Figure \ref{fig:comparison_profiles}. 
We observe that while there are slight differences in the extent of the convective envelope resulting from varying background conditions, 
these variations introduce only minor changes to $R_\star$, 
and $\dot{M}_{\rm g}$ as given by Equation \ref{eqn:Mdotgravitylimit} remains applicable.
Also consistent with the lower panel of Figure \ref{fig:simple_parameter_space}, 
the critical radius (the largest radius where $\langle v_R^2 \rangle_\rho$ crosses $\langle c_s^2 \rangle_\rho$ in the third panel of 
Figure \ref{fig:comparison_profiles}) first slightly shrinks for increasing density at fixed temperature $6\times 10^4$K, 
then expands again as temperature decreases with a fixed density of $10^{-10}$g/cm$^{-3}$.

For the pair of $\rho_{\rm out} = 10^{-10}$g/cm$^{-3}$ runs, 
and especially for $T_{\rm out}=3\times 10^4$K, 
the gas begins to exert some influence on the radiation field, 
leading to a temperature gradient from $R_\star$ towards the outer boundary instead of being completely isothermal.
This temperature distribution
should reduce the impact of enthalpy feedback as the gradient in $T_{\rm out}^4/\rho_{\rm out}$ diminishes, 
effectively reducing $f_2$ in a non-linear fashion. However, 
$c_{\rm s, rad, out}$ in this regime is already sufficiently low, 
such that gravitational feedback plays a more dominant role in determining $\dot{M}$ and $R_{\rm crit}$. Consequently, 
the outcomes remain consistent with our scalings.

\subsection{Effect of $\kappa(\rho_{\rm out}, T_{\rm out})$}
\label{sec:effect_kappaout}


In the above construction of the basic scaling laws, 
we have thus far made all comparisons regarding the opacity 
at $R_{\rm crit}$ as constant, 
assuming that the Eddington luminosity remains unaffected by environmental variables outside $R_\star$. 
However, 
$\langle \kappa \rangle_{L}$ may vary from the electron scattering opacity even outside $R_\star$ due to variation in temperature and density, 
introducing an overall change in $\lambda_{\rm diff}$. 
We consider below the validity of this approximation in the limit $\kappa(\rho_{\rm out}, T_{\rm out}) \neq \kappa_{\rm e}$.

\begin{figure}
    \centering
    \includegraphics[width=0.45\textwidth]{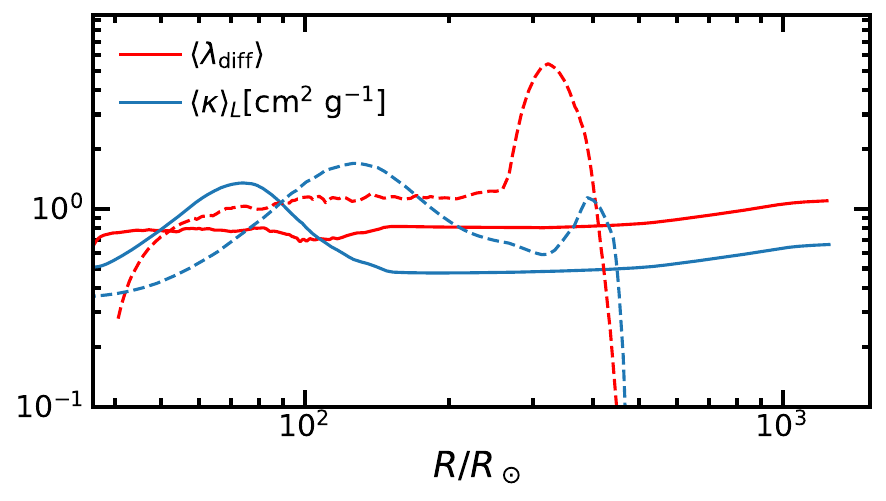}
    \caption{Similar to the lower panel of Figure \ref{fig:fiducial_1Denergyprofiles} but for run \texttt{D1e-10T3e4} (solid line) and \texttt{D1e-10T6e3} (dashed line). In the former case, we highlight that a large $\kappa(\rho_{\rm out}, T_{\rm out})$ (blue) leading to $\lambda_{\rm diff} \approx 1$ (red) at the outer boundary may not necessarily block accretion since $\kappa$ can approach the electron scattering opacity within a closer-in critical radius. 
    In the latter case, an extremely low $\kappa(\rho_{\rm out}, T_{\rm out})$ might lead to very strong accretion established at the outer regions, 
    which forms a shock at the some boundary where opacity suddenly increases and $\lambda_{\rm diff}$ becomes larger than unity. }
\label{fig:effect_of_kappa}
\end{figure}

In the case $\kappa(\rho_{\rm out}, T_{\rm out}) > \kappa_{\rm e}$, 
we can return to run \texttt{D1e-10T3e4} as an example. 
In Figure \ref{fig:effect_of_kappa} we plot the Eddington factor and opacity profiles from this run in solid lines.
At the outer boundary $\kappa(\rho_{\rm out}, T_{\rm out})$ 
is about twice as much as electron scattering 
due to the mild opacity bump around $10^4$ K. 
For this region there will be minimal accretion since even intrinsic luminosity $L_\star > L_{\rm Edd}$ 
when we use $\kappa(\rho_{\rm out}, T_{\rm out})$ to calculate $L_{\rm Edd}$. 
However, 
since $\kappa$ is a sensitive function of density and temperature, 
a mild rise in temperature inwards (see dashed lines in Figure \ref{fig:comparison_profiles}) relaxes $\kappa$ towards $\kappa_{\rm e}$ within a few hundred $ R_\odot$, 
where the super-sonic accretion flow 
is consistently established. 
Therefore, 
the true measurement of accretion rate is not too far off from using $\kappa_{\rm e}$ 
instead of using $\kappa(\rho_{\rm out}, T_{\rm out})$ in Equation \ref{eqn:Mdotgravitylimit}.

\begin{figure}
    \centering
    \includegraphics[width=0.48\textwidth]{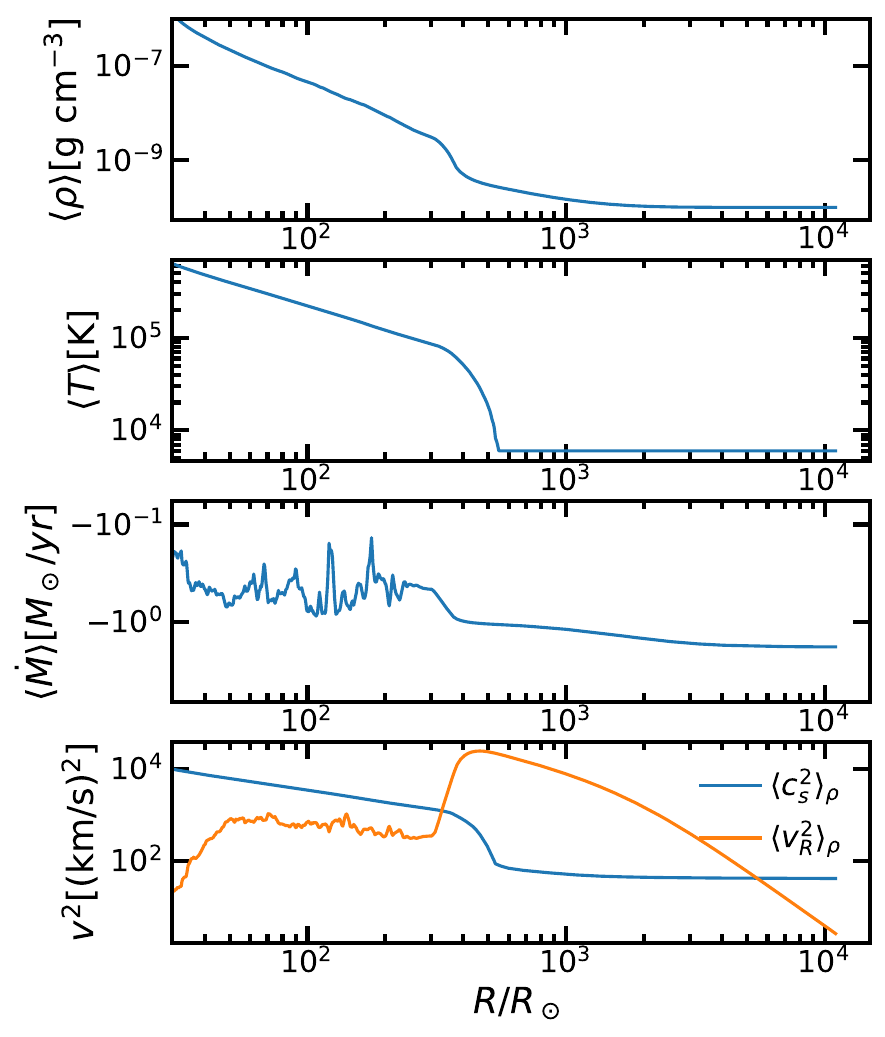}
    \caption{Similar to Figure \ref{fig:fiducial_1Dprofiles} but for run \texttt{D1e-10T6e3}. 
    The region around $\sim 500 R_\odot$ resembles a shock 
    where the supersonic accretion flow is stalled by local super-Eddington luminosity as opacity suddenly increases. To maintain a continuous mass accretion rate, 
    the transition in infall velocity is accompanied by a jump in density.}
\label{fig:lowkappa_1Dprofiles}
\end{figure}

In the other limit, dust sublimation leads to an opacity window, with 
$\kappa(\rho_{\rm out}, T_{\rm out}) \ll \kappa_{\rm e}$, over a
narrow and sensitive range of temperature around a few thousand $K$. 
However, contrary to a large $\lambda_{\rm diff}$, 
a negligible $\lambda_{\rm diff}$ corresponds to a large accretion rate established at a large $R_{\rm crit}$, close to the classical Bondi radius. 
If this radius is far enough from $R_\star$, 
the supersonic accretion flow may consistently persist inwards
until it reaches the region where $\kappa$ becomes significantly higher. 
To illustrate this point, we performed run \texttt{D1e-10T6e3} with negligible opacity at the outer boundary. 
The simulation domain of this run is set to $10^4 R_\odot$. 
The average 1D profiles over 100 orbits are plotted in Figure \ref{fig:lowkappa_1Dprofiles}.

We clearly observe that this particular run stands out from others, 
distinguished by significantly larger critical radius and accretion rate. 
Additionally, 
there is a notable density jump or infall velocity drop occurring at $\sim 400 R_\odot$. 
In Figure \ref{fig:effect_of_kappa} we also plot the local Eddington factor and opacity from run \texttt{D1e-10T6e3} with dashed lines to show that 
this mild shock structure is consistently associated with where 
$\kappa$ increases towards the electron scattering opacity and $\lambda_{\rm diff}$ abruptly rises to above unity. 
Consistency of Equation \ref{eqn:momentum} 
across the transition is only maintained through a change of sign in 
$c_s^2 - v_R^2$, 
and the steep change of $v_R$ also accompanies a steep change in $\rho$.
Consequently, 
a large accretion rate is allowed in the outer region due to low $\kappa(\rho_{\rm out}, T_{\rm out})$, 
while the inner region of high opacity within $\sim 400 R_\odot$ with sub-sonic infall velocity becomes 
an effective stellar envelope 
where this large accretion rate is 
deposited.

However, 
such high accretion rate is only quasi-steady for a short time. 
As its density rises from the accumulation of the 
accreted materials, the envelope's self-gravity becomes 
important on the mass-doubling timescale. 
It is unclear what consequences this pile up would have 
on the evolution of the stellar interior and $L_\star$. 
This issue is beyond the capability of our simulation
and the scope of this investigation. 

In addition, we emphasize, in the context of
realistic astrophysical settings such as an AGN disk, 
geometric effects becomes important 
in the limit that $R_{\rm crit}$, without the feedback effect,  
is significantly larger than the disk's vertical scale height $H$. 
In this region, departures of the ambient medium from spherical symmetry
distribution would invalidate the angular boundary conditions 
(\S\ref{sec:boundary}) and angular averages (Equation \ref{eq:averaging})
that we have adopted. 
The anisotropic mass and energy fluxes
may differ significantly from those presented here for 
isotropic-accretion flow  (see \S \ref{sec:implications}).

\subsection{The Slow Diffusion Limit}
\label{sec:slowdiffuse}

The self-consistency of the reduced-gravity approach lies on the fast-diffusion assumption, 
i.e. the characteristic infall velocity  
($V_{\rm R} (R_{\rm crit})$, which tends to be gas sound speed $c_{\rm s}$ in the fast-diffusion limit) is slower than photon diffusion velocity $c/\tau$. 
However, 
as we raise $\rho_{\rm out}$ towards higher values, 
we are bound to enter the very optically thick regime for which 
the preceding calculations are no longer applicable. 
This transition leads to not only the adiabatic expansion of the envelope 
as cooling becomes inefficient but also the coupling between the  
radiation and gas sound speed.

\begin{figure}
    \centering
\includegraphics[width=0.5\textwidth]{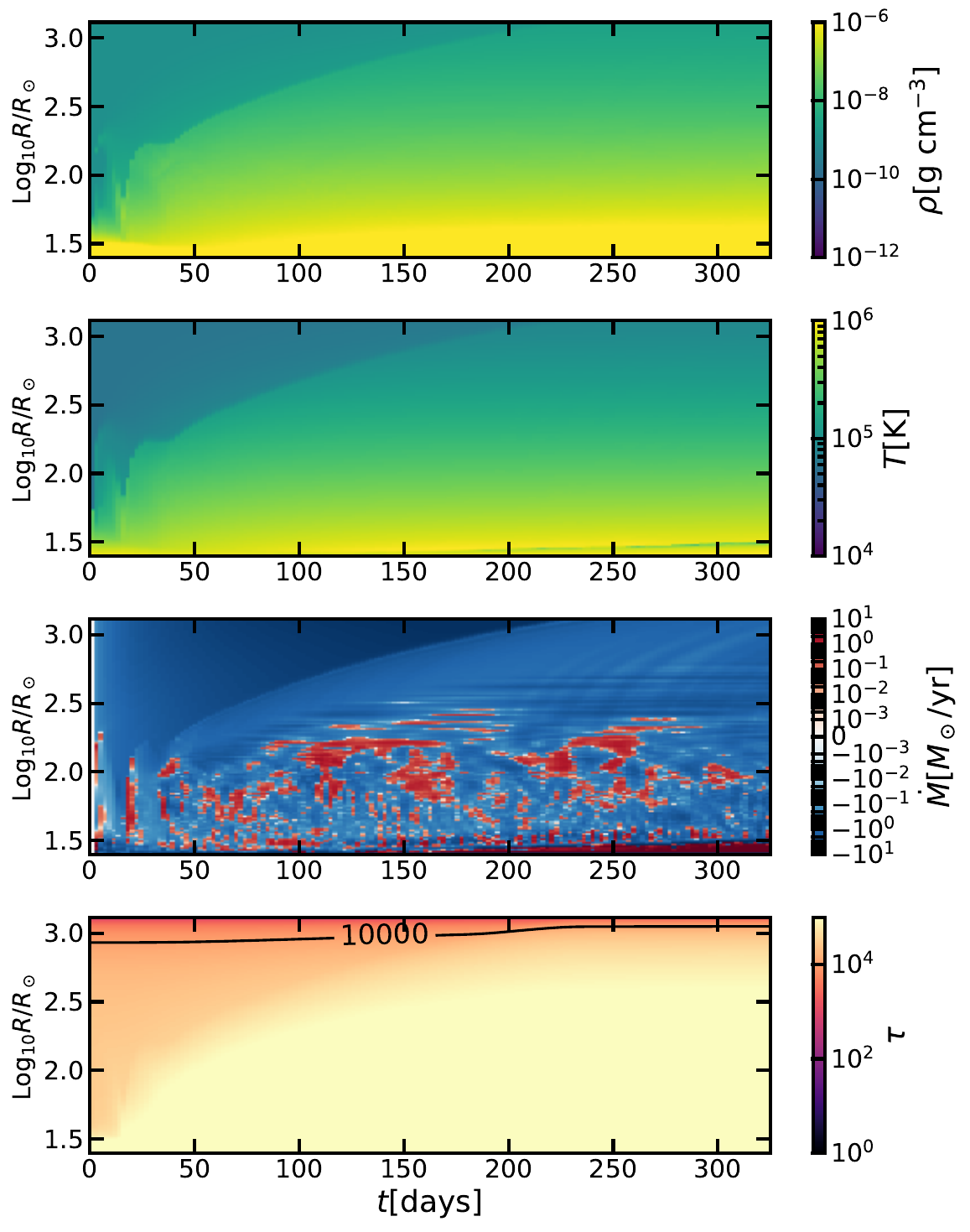}
    \caption{Angle averaged radial distributions of density, temperature, radial accretion rate and optical depth as functions of time (horizontal axis) for the run \texttt{D1e-10T6e3}.}
\label{fig:spacetime_adi}
\end{figure}

As an example, 
Figure \ref{fig:spacetime_adi} shows the spacetime plot for run \texttt{D1e-9T6e4}. 
Since the optical depth in the convective zone is large enough $\tau > c/c_{\rm s} \sim 10^4$, the 
diffusive luminosity can no longer be modeled simply as a reduction in gravity. 
The initial fast accretion of the system resembles the Bondi accretion process of a mixture of radiation and gas with sound speed $c_{\rm s, tot}$ defined by total pressure.
But the accreted materials 
immediately form a shock in the inner envelope where pressure is high enough to overcome the ram pressure of the accretion flow. 
The post-shock material accumulates and becomes hydrostatic (with low, subsonic fluctuations in $v_R$). 
In some sense this expansion is similar to \texttt{D1e-10T6e3} in \S \ref{sec:effect_kappaout}, 
the key difference being that the material infall outside the shock is faster than the sound speed defined by \textit{combined} pressure of radiation and gas, rather than the significantly lower gas-only sound speed in the fast-diffusion limit. 
We show below that this expansion phase, as well as the long-term (on the thermal timescale) evolution of run \texttt{D1e-9T6e4} can be well-reproduced and traced by an adiabatic simulation 
without radiative transfer.

\subsubsection{Long term evolution in Adiabatic simulations}

\begin{figure}
    \centering
    \includegraphics[width=0.46\textwidth]{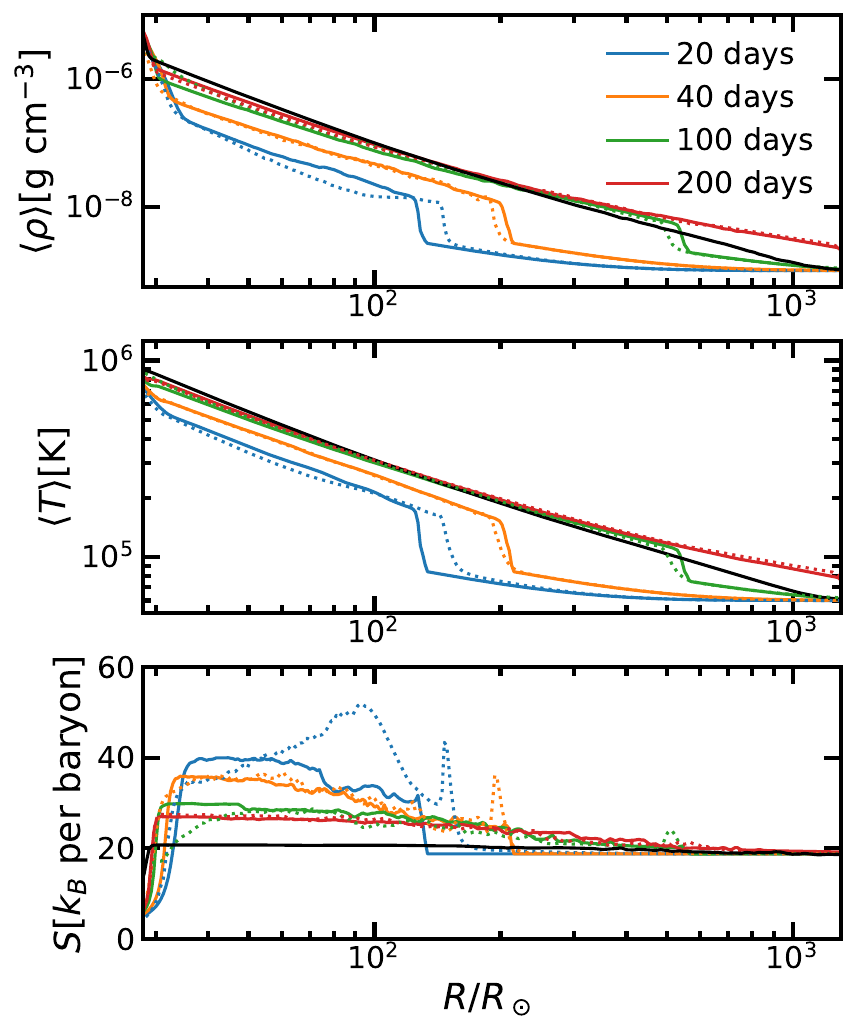}
    \caption{The angle-averaged radial distributions of 
    $\langle \rho \rangle, \langle T \rangle$ as well as entropy from run \texttt{D1e-9T6e4} (dotted lines) and the corresponding adiabatic run \texttt{D1e-9T6e4\_adi} (solid lines) without radiative transfer at different snapshots during the expansion phase. 
    We also plot the long-term average profiles from run \texttt{D1e-9T6e4\_adi} in solid black lines after it evolves into an oscillating quasi-steady state (see Figure \ref{fig:spacetime_adi_woRT}).
    The final profile is close to a hydrostatic structure, with radial-velocity fluctuations much lower than the combined sound speed of radiation and gas pressure.}
\label{fig:adi_1Dprofiles}
\end{figure}

We perform adiabatic simulation \texttt{D1e-9T6e4\_adi} 
by turning off radiative transfer (RT) and 
using a modified equation of state 
(containing radiation and gas at the same temperature), 
and run for 10 times longer than \texttt{D1e-9T6e4} to cover a larger number of dynamical cycles. 
In Figure \ref{fig:adi_1Dprofiles} 
we plot the 
density and temperature evolution of the initial expansion phase for both  \texttt{D1e-9T6e4} (dotted lines) and \texttt{D1e-9T6e4\_adi} (solid lines), confirming that the evolutions are very similar. We also plot the evolution of specific entropy per baryon in the lowest panel of Figure \ref{fig:adi_1Dprofiles} for these two simulations, which is defined as 
\begin{equation}
    {S \over k_B} = \dfrac{4 P_{\rm rad}}{P_{\rm gas}} + \dfrac12 \ln \left(\dfrac{P_{\rm rad}}{P_{\rm gas} \tilde{\rho}}\right), 
    \label{eqn:entropy_mixture}
\end{equation}
for a mixture of gas and radiation at the same temperature \citep{GoodmanTan2004} and constant molecular weight.
Here, $\tilde\rho=\rho/\rho_0$ with $\rho_0$ being some constant reference density.  The choice of $\rho_0$ affects only an unimportant additive constant in the entropy and, for convenience, we set it to unity in code units.
In the first 100 days, we see that the post shock material is heated to a higher entropy than the equi-entropy accretion flow 
from the outer boundary, but gradually its entropy converges to $S (\rho_{\rm out}, T_{\rm out})$ after the shock/envelope propagates to the outer boundary.

\begin{figure}
    \centering
\includegraphics[width=0.5\textwidth]{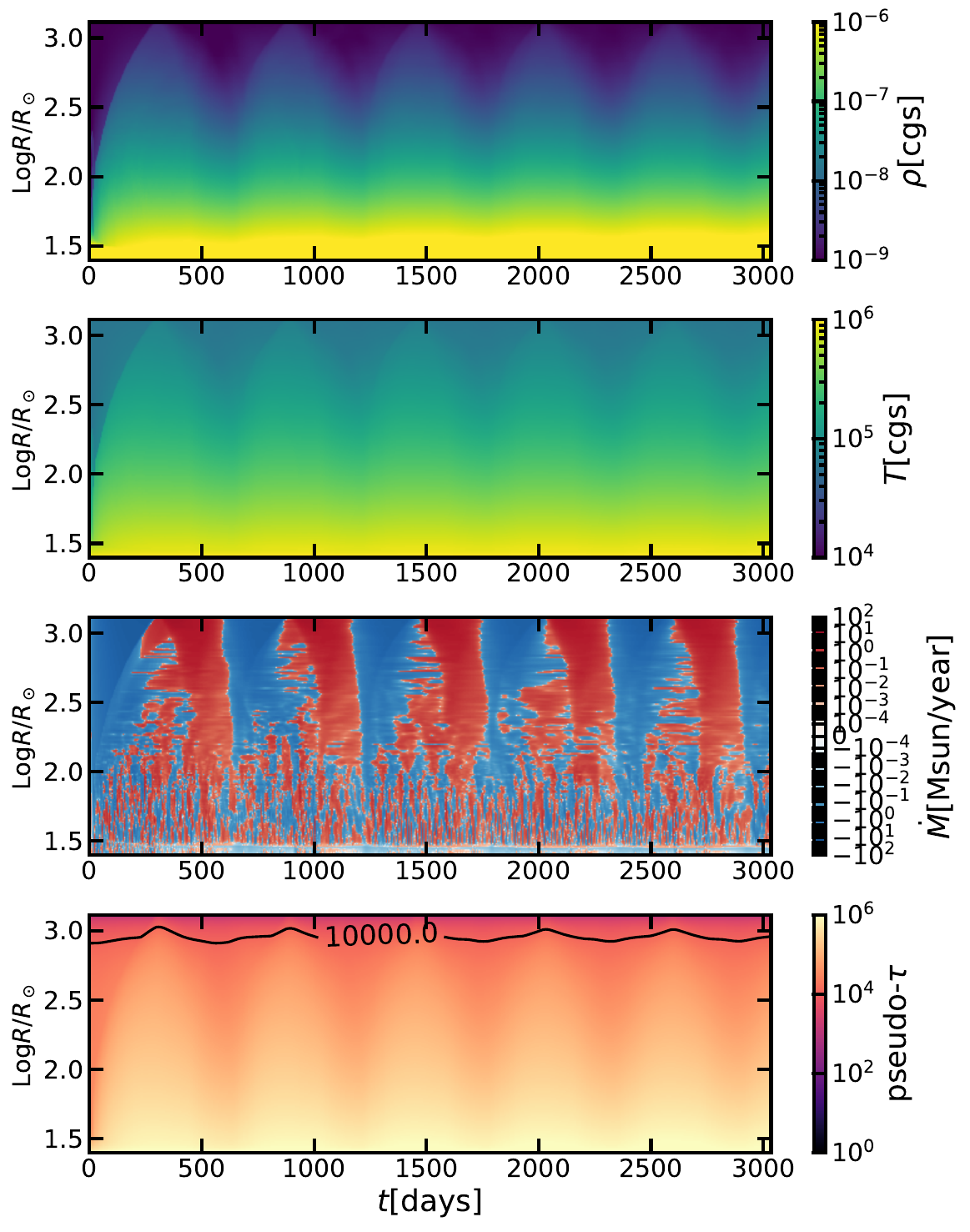}
    \caption{Angle-averaged radial distributions of density, temperature, radial accretion rate and optical depth as functions of time (horizontal axis) for the adiabatic run with modified equation of state \texttt{D1e-9T6e4\_adi}. 
    The bottom panel shows the nominal electron-scattering optical depth integrated from the outer boundary (although radiative transfer is not actively modeled in this run).}
\label{fig:spacetime_adi_woRT}
\end{figure}

When the shock propagates to $R_{\rm out}$ 
at a timescale of $R_{\rm out}/c_{\rm s, tot, out}$, 
the total mass of materials contained in this hydrostatic, isentropic envelope is a few solar mass. 
We have performed another run which contains a radial-self-gravity treatment \citep[similar to][]{Goldberg2022}
and confirm this marginal self-gravity does not make any significant difference to the expansion of the envelope. 
We extend the inexpensive adiabatic simulation \texttt{D1e-9T6e4\_adi} over multiple dynamical timescales to confirm that isentropic profile is maintained after time-average, 
which is plotted in Figure  \ref{fig:adi_1Dprofiles} using solid black lines.
The system does not evolve into a complete steady state. 
Instead, 
inflow-outflow oscillations still occur on the largest dynamical timescale beyond $\sim 100 R_\odot$, 
possibly because the domain is mildly convective for a slightly negative entropy gradient, 
as shown in the spacetime plot of Figure \ref{fig:spacetime_adi_woRT}. 
This does not markedly affect the time-average profile because the time-integrated net mass flux over many cycles is small compared to the mass accumulated within the simulation domain during the first expansion phase.

\begin{figure}
    \centering
\includegraphics[width=0.45\textwidth]{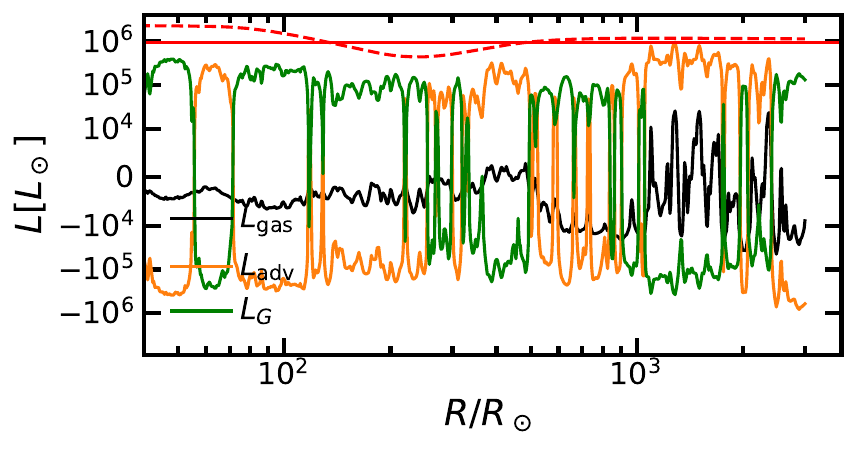}
    \caption{Average radial profiles for different energy fluxes in the adiabatic run \texttt{D1e-9T6e4\_adi}, 
    compared to the initial $L_\star$ (red solid line) or the diffusive luminosity post-processed from the average temperature and density profiles (red dashed line), which could probably act as a stable net cooling term \textit{if} radiative transfer is turned on. 
    The fact that these advective luminosities become $\ll L_*$ suggests that the system is approaching a hydrostatic state, 
    and that further accretion would occur on a thermal rather than free-fall timescale 
    (if radiative transfer were fully taken into account).}
\label{fig:longterm_energy}
\end{figure}

Regarding the long-term evolution of the envelope, we compare the energy fluxes for run \texttt{D1e-9T6e4\_adi}, 
averaged over the last 2000 days, 
with the initial stellar luminosity $L_\star$ (red solid line, although this quantity is not actively modeled in the purely adiabatic run) in Figure \ref{fig:longterm_energy}. 
Despite the strong positive-negative fluctuation of energy fluxes, 
associated with the mass inflow-outflow oscillations, the net advective 
energy transport, averaged over many cycles, 
become substantially lower than $L_{\star}$. We also plot the diffusive luminosity profile post-processed from the equilibrium temperature and density profile (applying $F_{\rm diff} = - \frac{4 a c}{3 \kappa (\rho, T) \rho} \frac{ T^3 \mathrm{d} T}{\mathrm{~d} r}$) in red dashed lines, 
which is no longer a constant of radius but still similar to $L_\star$ and dominate over other average luminosities. This hierarchy implies
that the envelope would probably contract on a Kelvin-Helmholtz (KH) 
timescale $L_{\rm diff}R_\star/GM_\star M_{\rm env}$ if the effect of 
radiative transfer is taken into account. Such full-scale RT simulations 
would require computational resources beyond the present availability.

\subsubsection{Self-gravity of an Isentropic Envelope}

Extending our analysis to a broader range of parameters, 
it is useful to provide some further examination 
on the general conditions for the hydrostatic envelope to 
become self-gravitating.

During the advanced stage ($ t \gtrsim 10^3$ days) of the adiabatic 
simulation, radiation pressure dominates ($P_{\rm rad} \gtrsim 
P_{\rm gas}$) with slowly varying adiabatic index ($\sim 4/3$) 
in the isentropic envelope (Figure \ref{fig:spacetime_adi_woRT}).
In this case, most of the envelope's mass $M_{\rm env}$ is contained 
in the outer envelope (Figure \ref{fig:adi_1Dprofiles}).  
For the outer boundary of the 
material that contributes to the gravitational field,
we adopt the Bondi radius defined 
by total sound speed in the slow-diffusion limit, 
\begin{equation}
R_{\rm B, tot} = \dfrac{GM_\star}{2c_{\rm s, tot, out}^2}.
\end{equation}

and therefore 

\begin{equation}
    M_{\rm env} \sim 4\pi \rho_{\rm out} R_{\rm B, tot}^3 \sim 4 \pi\left(G M_*\right)^3 \frac{\rho_{\rm out}}{8 c_{\mathrm{s, tot, out}}^6}.
    \label{eqn:Menv}
\end{equation}

For a calibration, we measure, from the results in 
run \texttt{D1e-9T6e4}, the mass within $R_{\rm B, tot} \approx 1047 R_\odot \lesssim R_{\rm out}$ to be $ M_{\rm env} \sim 4 M_\odot$, 
which is in the regime where self-gravity is insignificant 
and consistent with the above analytic expectation. 
However, Equation \ref{eqn:Menv} indicates $M_{\rm env}/M_\star$ 
to be a rapidly increasing function of $M_\star$.  
For the Eddington model with a constant 
$\beta_\star = (1 + P_{\rm rad, \star}/P_{\rm gas, \star})^{-1}$,
the structure of massive stars is well approximated by a $n=3$ 
polytrope.  In this model, the interior mass is related to its 
average gas pressure fraction $\beta_\star$ and molecular weight 
$\mu_\star$ by \citep[e.g.][]{GoodmanTan2004}

\begin{equation}
    M_*^2 \approx \frac{1}{(0.3639 G)^3}\left(\frac{3}{a}\right)\left(\frac{k_{\mathrm{B}}}{\mu_\star m_p}\right)^4 
    \frac{1-\beta_*}{\beta_*^4}.
    \label{eq:Mstarout}
\end{equation}

Combining Equations \ref{eqn:Menv}, \ref{eq:Mstarout} with
the environment parameters in terms of $\beta_{\rm out} = (1 + P_{\rm rad, out}/P_{\rm gas, out})^{-1}$, we find

\begin{equation*}
    M_{\rm env}/M_* \sim \left(\frac{\mu}{\mu_*}\right)^4 \times \frac{1-\beta_*}{\beta_*^4} \times \frac{\beta_{\rm out}^4}{1-\beta_{\rm out}}.
\end{equation*}
In the above expression, 
we distinguish between the mean molecular weight of the star $\mu_\star$ and 
that of surrounding materials $\mu$.  When $\mu \sim \mu_\star$, 
the envelope mass fraction is determined almost entirely by the star's mean gas-pressure fraction relative to that of the ambient medium. 
For the same set of environment parameters, even a low mass-envelope 
can transition to a self-gravitating state as $\beta_\star$ decreases 
with the growth of stellar mass and initiate runaway accretion. In
such a scenario, a hydrostatic equilibrium may not be established 
until the star gain sufficient mass to open a gap in the disk
\citep{Lin1993}.
This transition is also about the crossover point where the radiation entropy 
of the background material would become close to that on the star's surface 
(Equation \ref{eqn:entropy_mixture}).

\section{Accretion Rates and Critical Radii in Realistic Disk Environments}
\label{sec:implications}

\begin{figure}
    \centering
    \includegraphics[width=0.48\textwidth]{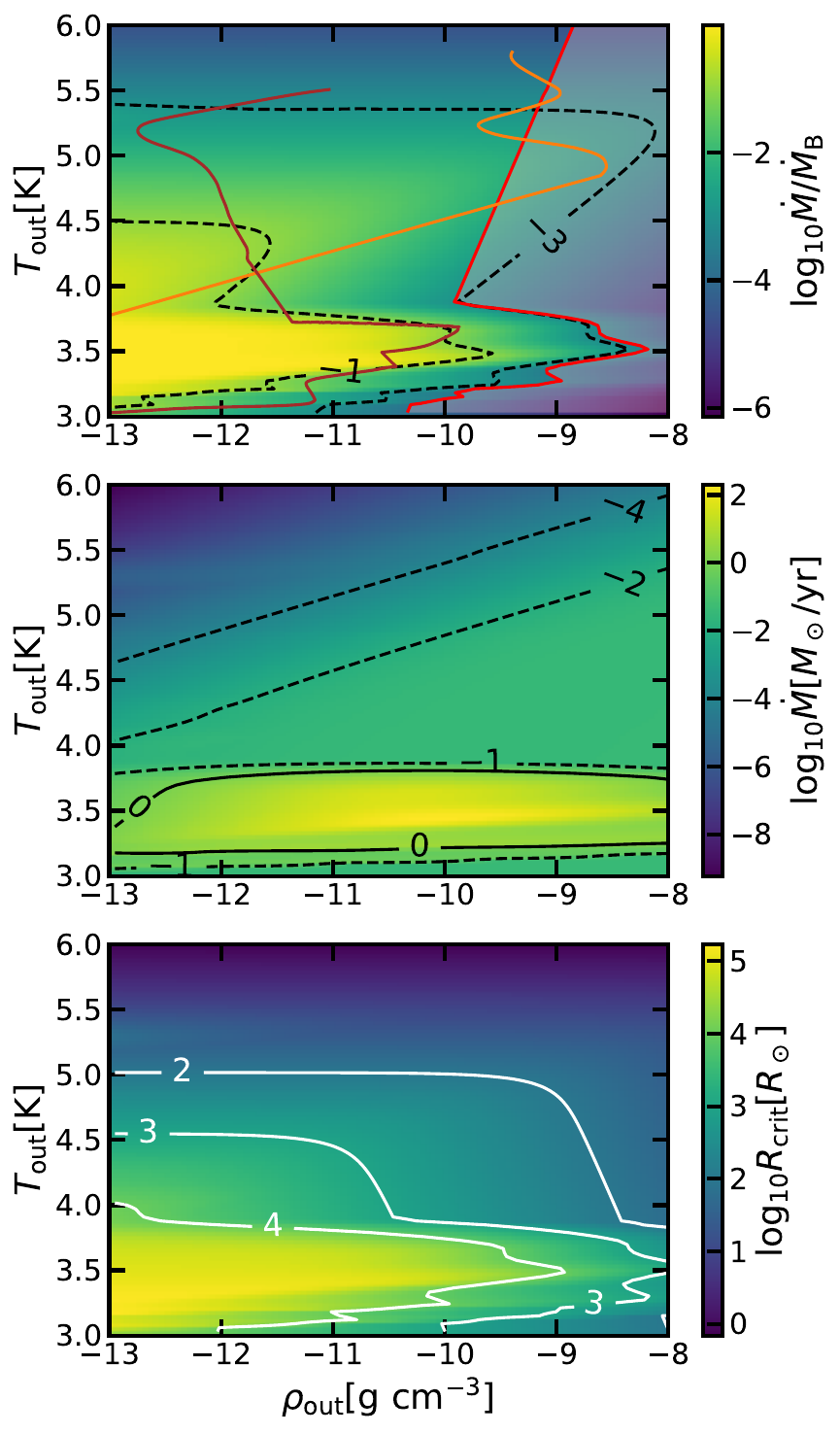}
    \caption{Solutions to 
    Equation \ref{eqn:quartic_for_Mdot} as functions of $T_{\rm out}$ and $\rho_{\rm out}$. 
    Similar to Figure \ref{fig:simple_parameter_space} but taking the opacity at critical radius to be the smaller  of $\kappa_{\rm e}$ and $\kappa(\rho_{\rm out}, T_{\rm out})$ to obtain upper limits for the accretion rate. The optically thick regime to the right of the red line ($\rho_{\rm out} \kappa_{\rm crit} R_{\rm B} \sim  c/c_{\rm s, out}$) is highlighted. The typical \citet{Sirko2003} and \citet{Thompson2005} disk models are plotted in orange and brown solid lines in the top panel, respectively.}
\label{fig:comp_parameter_space}
\end{figure}

Based on the results in \S\ref{sec:effect_kappaout},
we apply $\kappa_{\rm crit} = \min[\kappa(\rho_{\rm out}, T_{\rm out}), \kappa_{\rm e}]$ 
in Equation \ref{eqn:quartic_for_Mdot} to obtain upper limits for stellar accretion rate. 
Figure \ref{fig:comp_parameter_space} 
is a modified version of Figure \ref{fig:simple_parameter_space} with self-consistent opacity, 
although we also highlight the adiabatic regime where the reduced-gravity approach no longer applies (right of the red lines). 
Similar to the discussion in \S \ref{sec:feedback} and \ref{sec:verify_scaling},
we observe that for $T_{\rm out} \lesssim 10^4$K, 
very high Bondi-like accretion rate might be allowed despite 
the scalings in other parameter space.

As an application of our scaling to realistic AGN environments, 
we generate two typical 
\citet[][hereafter SG]{Sirko2003} and \citet[][hereafter TQM]{Thompson2005} 
models for accretion disk around a $10^8 M_\odot$ supermassive black hole (SMBH).
We adopt the default setup of the AGN disk modeling tool 
\texttt{pAGN} \citep[][other default parameters see documentation 
therein]{Gangardt2024}. The SG model extends and connects, inside-out, a standard disk solution \citep{SS1973} with that for a self-gravitating (Toomre $Q=1$) disk.
In this prescription, the inward velocity $v_{\rm acc}$ of gas in the accretion disk 
is derived from the viscous stress \citep{LyndenBell1974} driven by turbulent viscosity such that 
$v_{\rm acc} \propto \alpha c_{\rm s, tot, out}^2$ (or simply $c_{\rm s, rad, out}^2$ when 
dominated by radiation pressure) with $\alpha\sim 0.01$.  

The TQM model is constructed under the assumption of trans-sonic 
radial velocity, i.e. $v_{\rm acc} \propto m c_{\rm s, tot, out}$ with $m\sim 0.2$
rather than that through viscous diffusion.  Moreover, it takes into account the
reduction in the mass flux due to the continuous star formation enroute 
the inward accretion flow from the outer disk regions towards the SMBH, 
For a given mass flux at the outer regions of AGN disks, the TQM model 
would imply significantly lower densities for the gas reaching the 
high-temperature inner regions than that derived with the SG model.

We plot the ${\dot M}$ and $R_{\rm crit}$ contours in the 
$\rho_{\rm out}, T_{\rm out}$ parameter plane for the
SG (orange lines) and TQM (brown lines) models in 
Figure \ref{fig:comp_parameter_space}.  
These solutions traverse through different regimes 
of parameter space. In most of the star-forming outer ($R_{\rm AGN} 
\gtrsim 10^{-2}$ pc) regions of the disk where $T_{\rm out}$ 
and $c_{\rm s. rad, out}$
are relatively low, gravitational feedback sets the upper limit 
for the stellar accretion rate ${\dot M}_\star$. 
Although star formation may be quenched 
in the high-$\rho_{\rm out}$, high-$T_{\rm out}$, stable ($Q \geq 1$)  
inner ($R_{\rm AGN} \lesssim 10^{-2}$ pc) regions of the disk, 
these locations may still 
contain captured or inwardly-migrated stars.  Here, the 
environment inferred by the SG model leads to the 
adiabatic-accretion regime, whereas,
due to much lower densities, that inferred with the TQM model 
enters the enthalpy feedback regime.  
There is also a particular region 
in the TQM model where, due to the very low opacity at $T_{\rm out}
\sim 10^3$K, rapid/dynamical accretion might be allowed.

\begin{figure}
    \centering
    \includegraphics[width=0.45\textwidth]{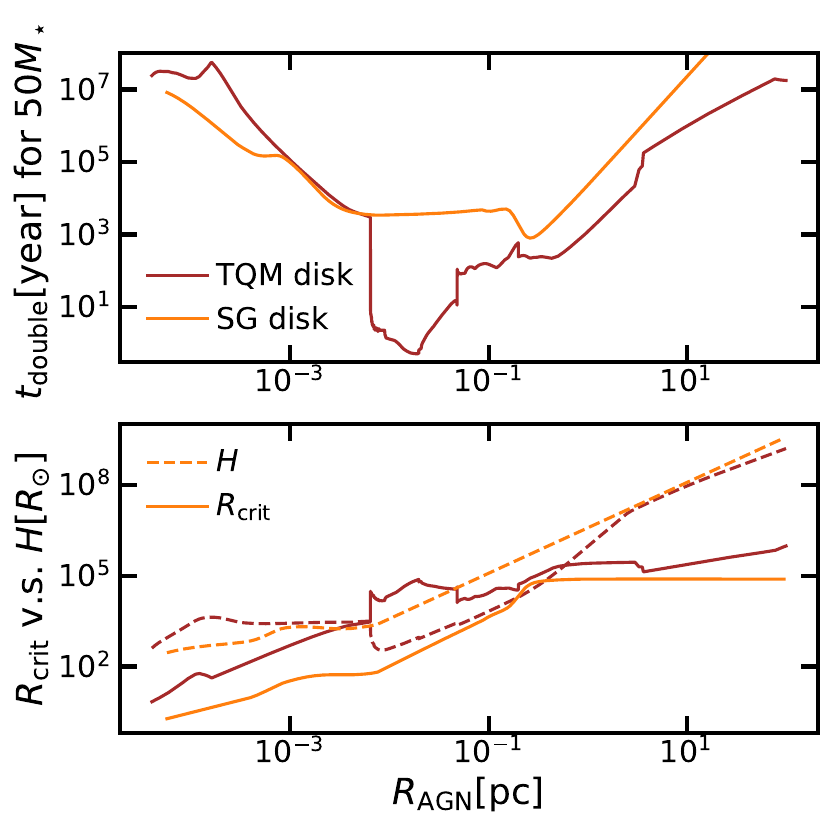}
    \caption{Top panel: 
    Mass doubling timescale for a 50$M_\odot$ star measured in years 
    as a function of distance to the center supermassive black hole in typical AGN disk models. 
    Lower panel: critical radius assuming isotropic accretion (solid lines) versus the disk scale height (dashed lines) in AGN disk models. Anisotropic and rotational effects are expected to be important for $R_{\rm crit} \gtrsim H$. These calculations assume fast radiative diffusion, which might not apply in the densest region of the SG disk.}
\label{fig:tdouble_and_Rcrit_for_disks}
\end{figure}

Under the fast-diffusion assumption, we evaluate
the doubling timescale $t_{\rm double} = M_\star/\dot{M}$ 
for a 50 $M_\odot$ star across the AGN disk for the 
SG and TQM models (the top panel of Figure 
\ref{fig:tdouble_and_Rcrit_for_disks}).  
We note that the doubling timescale 
is significantly prolonged ($t_{\rm double} \gtrsim 1$ Myr) close to the inner 
$R_{\rm AGN} \lesssim$ a few $10^{-4}$ pc) and outer ($R_{\rm AGN} \gtrsim $ 
a few pc) boundaries of the accretion disk, while, in regions around 
$0.001-0.1$pc, $t_{\rm double} \sim 10^4$ yr in the SG model
and $\lesssim 10^2$ yr in the TQM model.

Since $\dot{M_{\rm g}}/\dot{M_{\rm r}} \propto M_\star$,
the gravitational-potential feedback limit ultimately 
dominates as the star's mass increases. For high-mass stars,
it is especially important to note that the accretion rate
only depends on $(1-\lambda_\star) R_\star$ rather than 
the mass of the star (see Equation \ref{eqn:Mdotgravitylimit}). 
If $(1-\lambda_\star) R_\star$ decreases with $M_\star$ 
(due to the rising $\lambda_\star$), the asymptotic 
doubling time $t_{\rm double}$ would increase with $M_\star$,
instead of decreasing with $M_\star$ as in the classical 
Bondi-accretion model.  Consequently, the accretion process 
would self-limit the asymptotic value of $M_\star$. 

In addition to the non-linear feedback from the 
stellar evolution itself that might arise in this process, 
it's crucial to evaluate the importance of the disk geometry,
i.e. an anisotropic $\rho_{\rm out}, T_{\rm out}$ boundary condition by comparing 
$R_{\rm crit}$ with $H$ throughout $R_{\rm AGN}$. Adopting the 
fast-diffusion approximation, the profiles of 
$R_{\rm crit}$ (solid lines) and $H$ (dashed lines) are plotted 
for both the SG (brown) and TQM (orange) models (the 
lower panel of Figure \ref{fig:tdouble_and_Rcrit_for_disks}).
Under effect of strong thermal feedback, 
the modified $R_{\rm crit} \lesssim H$ so that the isotropic-accretion
assumption is fully satisfied for the SG model and partially satisfied
(except in the $R_{\rm AGN} \sim 0.01-1$ pc region, see below) 
for the TQM model in most of the relevant parameter space.

In the fast diffusion regime, the gravitational-feedback limit eventually dominates as the star 
grows in mass. In this regime, the critical radius approaches 

\begin{equation}
\begin{aligned}
    &R_{\rm crit, grav} =  \sqrt{\dfrac{(1-\lambda_\star) R_\star c}{\rho c_s \kappa}} \\& <  2\times 10^4 R_\odot\left(\dfrac{R_\star}{R_\odot}\right)^{0.5}  \left(\dfrac{\rho}{10^{-16} {\rm g cm}^{-3}} \dfrac{c_s}{10^7 {\rm cm/s}} \dfrac{\kappa}{{\rm cm}^{2} {\rm g}^{-1}}\right)^{-0.5},
    \end{aligned}
    \label{eqn:rcrit_grav}
\end{equation}

which in the SG model is $\ll H$ even in  the very low-density outskirts 
of AGN disks regardless of the stellar mass. When $L_* \approx  L_\mathrm{Edd}$
(and $\lambda_\star \rightarrow 1$), the true critical radius can further reduce
to much smaller values than this nominal upper limit.

The low-opacity regime of the TQM model (at $R_{\rm AGN} 
\sim 0.01-1$ pc where $T_{\rm out} \sim 10^3$K, top panel of 
Figure \ref{fig:comp_parameter_space}) is particularly 
special, since it is the only location where a large 
$R_{\rm crit}$ is both close to the classical Bondi radius 
and $\gtrsim H$ (Figure \ref{fig:tdouble_and_Rcrit_for_disks}) due to minimal radiative feedback. 
It is uncertain how anisotropic effects such as 
density and temperature stratification as
well as background shear may influence the 
final accretion rate.   
Recent 3D simulations 
of analogous ``super-thermal'' planetary accretion
(with $R_{\rm Bondi} \geq H$) show a meridional 
flow pattern.  In the same plane as the global
protoplanetary disk, gas flows away from the planet
in a circumplanetary disk (CPD) which is nearly
centrifugally supported against the planet's gravity.
Above the CPD, gas also flows towards and being accreted 
onto the planet, with highest flux in the mid-declination  
\citep{Choksi2023,Li2023}.  This anisotropic flow
pattern leads to a different scaling (with respect to 
disk property and companion mass) relation for 
the net accretion rate. When the turbulent viscosity is 
high enough to overcome the potential-vorticity barriers
in the protostellar disk, the (companion-centric) 
azimuthal rotation in the CPD 
plays a minor role in determining the accretion rate 
scaling \citep{LiChen2021}. 
However, the density stratification normal to the AGN disk (over a 
characteristic scale height $H$) introduces highly anisotropic opacity 
profiles in different directions, which can complicate the determination 
of the accretion rate and an effective $R_{\rm crit}$ when it is 
$\gtrsim H$. 
For these reasons, more realistic shearing box
simulations need to be conducted in the future to investigate the 
effect of geometry under these circumstances.
For the rest of the parameter space, 
our scalings derived from isotropic accretion is sufficient for 
more accurate modeling of AGN stars' long-term evolution, 
following the works of \citet{Cantiello2021,Dittmann2021,AliDib2023}. 

The fast-diffusion approximation is valid in most of the relevant parameter space.
Nevertheless, adiabatic accretion could occur in the region where $\rho_{\rm out} 
\approx 10^{-9} {\rm g/cm}^3$ (Figure \ref{fig:comp_parameter_space})
For the SG model, such high densities are reached at the marginally
self-gravitating $R_{\rm AGN} \sim$ 0.01-0.1 parsec. Under these circumstances, 
a hydrostatic envelope tends to form on top of the stellar convective surface 
and it extends to at least the Bondi radius defined by total (gas+radiation) sound 
speed (\S \ref{sec:slowdiffuse}). The accumulation of mass over time may cause this
region to become self-gravitating and expand the envelope beyond the 
disk scale height. 
In 1D simulations of the star's long-term evolution,  
extra care needs to be taken in modeling the cooling of such envelopes.


\section{Summary and Future Prospects} 
\label{sec:conclusions}

\begin{figure}
    \centering
    \includegraphics[width=0.45\textwidth]{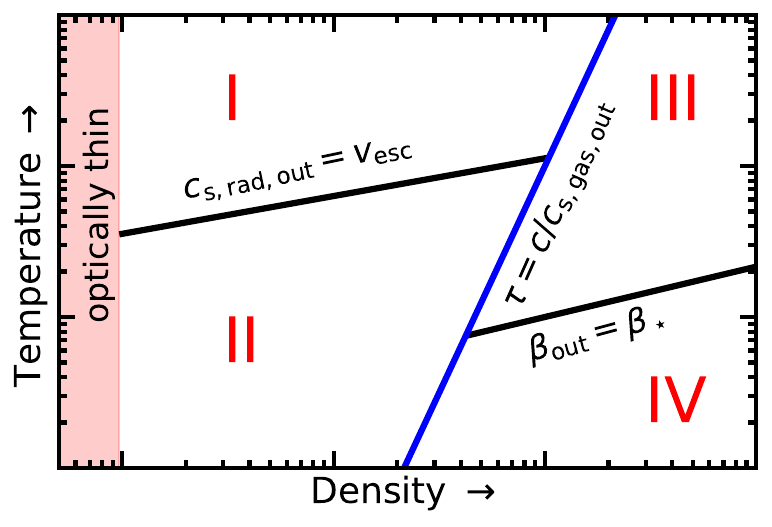}
    \caption{{Schematic for different accretion regimes we identify in this paper, and the criteria that separate them. I) fast diffusion, radiation enthalpy feedback dominated; II) fast diffusion, gravitational feedback dominated; III) slow diffusion, non-self-gravitating; IV) slow diffusion, self-gravitating. The optically thin regime is not explored in this paper.} }
\label{fig:schematic}
\end{figure}

In this paper, 
we performed radiation hydrodynamic simulations to study the accretion of stars in isotropic gas-rich environments. Across a wide range of parameter 
space for the background density and temperature, 
the outcome of accretion onto a given stellar envelope, is quite diverse and {can be summarized in the schematic Figure \ref{fig:schematic}.} 
{ To the left of the blue line ($c/\tau > c_s$), radiation diffuses outward more rapidly than the gas flows inward.
This ``fast-diffusion" region is further divided into two subregions according as the radiation sound speed at the outer boundary is greater (region I) or less (region II) than the escape velocity from the stellar surface.
In region II, accretion is throttled mainly by the conversion of gravitational to luminous energy at the stellar surface: $L_{\rm grav}=GM\dot M/R_*\lesssim L_\mathrm{Edd}$.
This is Eddington-limited accretion as usually calculated for accretion onto neutron stars or black holes \citep{Thorne1973,Thorne1974}, but at a higher absolute rate because $GM_*/R_*\ll c^2$.
In region I, radiative advection makes the accretion rate less than would be predicted by the classical Eddington limit.}

{
Nevertheless, we expect that if the flow were completely optically thin, the accretion rate should be determined by the Eddington limit regardless of the ordering of $v_\mathrm{esc}$ and $c_\mathrm{s,rad,out}$, since advection of the radiation could hardly occur; 
the distinction between regions I and II would then disappear sufficiently far to the left of the blue line (pink region).
However, all of our simulations are at least moderately optically thick ($\tau \gtrsim10^2$-$10^3$), so that the advection of radiation enthalpy cannot be entirely neglected. 
In practice, the true optically thin scenario is never encountered in our AGN context, therefore we do not delve into the specifics of this transition.}


In the high density, slow-diffusion limit ($c/\tau < c_s$), 
radiation and gas integrate into a combined fluid. 
Under this condition, the envelope structure tends to become hydrostatic 
and the net accretion rate is quenched。
{
In fact, the slow-diffusion regime is analogous to the well-studied photon trapping limit of black holes \citep{Begelman1978, Thorne1981, Flammang1982, Inayoshi2016, Begelman2017, Wang2021} where accretion rate can be hyper-Eddington. 
However, unlike the the even horizon of a black hole, 
the stellar surface exerts pressure that promotes the establishment of a hydrostatic envelope in the absence of self-gravity, resulting in the emergence of regime III. }

{
If the gas-pressure fraction of the ambient medium ($\beta_\mathrm{out}$) is greater than that of the star ($\beta_*$), the self-gravity of the envelope cannot be neglected.
In this regime (region IV), we expect runaway accretion on a dynamical ($\ll$ thermal) timescale.
Hydrostatic Eddington models are $n=3$ polytropes, with a monotonic relation between $M_*$ and $\beta_*$ (Equation \eqref{eq:Mstarout}); if one tries to increase the mass by adding material with $\beta<\beta_*$ quasi-adiabatically, there is no hydrostatic solution.
We have not modeled this regime in detail in the present paper.
However, stars that form in an AGN disk via self-gravity (i.e., via the Toomre instability rather than by capture or migration) could start at a mass such that $\beta_*\approx\beta_\mathrm{out}$, so that such stars would accrete in region IV, except insofar as their accretion rate is modified by angular momentum, gap opening, and mergers---all of these being effects that cannot be modeled by the spherically-symmetric, homogeneous ambient medium assumed here.}

We apply the scaling results obtained from our simulations
to a wide range of boundary conditions depicted by the
SG and TQM models of AGN disks over an extended radial scale, and found they typically accrete in the fast diffusion regime.
For the outer star-forming region, embedded stars' accretion rate 
is primarily limited by gravitational feedback (Eqn \eqref{eqn:Mdotgravitylimit},  region II).
Although {\it in situ} star formation is quenched by the 
local gravitational stability, embedded stars may still migrate
to or capture at the innermost region of the AGN disks. 
At this location, stars' infalling envelope is optically thick
and their accretion rate is mostly limited by the 
enthalpy-feedback effect (I).

In the intermediate disk radius, dust sublimation leads to an
opacity valley.  Consequently, the upper limit of accretion rate
can be close to the Bondi rate with a relatively short mass-doubling
timescale.  Moreover, the critical radius becomes comparable to or larger
than AGN-disk's density scale height normal to the disk plane,  in the 
TQM (but not the SG) model.  Such a boundary condition would invalidate
the isotropic assumption we have adopted for both the numerical simulation
and analytic approximation. This inconsistency warrants some follow-up 
investigations with shearing-box simulations to capture the
geometric effects for the anisotropic accretion. The disk thermal 
structure may also be significantly modified by sufficiently
intense total luminosity of $N$ coexisting stars with an area 
filling factor $N (R_{\rm crit}/R_{\rm AGN})^2 \sim {\mathcal{O}} (1)$.

Having laid out a framework of estimating the critical radius and 
accretion rates for moderately massive ($50 M_\odot$) stars, we 
will proceed, in parallel, to study more massive ($ \geq 10^2 M_\odot$)
stars to validate the mass dependence of the limiting accretion-rates.
These massive stars may emerge directly through gravitational instability
\citep{GoodmanTan2004, Jiang11, Chen2023} or in regions of the disk
where the mass doubling time is very short.
We will also investigate the competition between possible optically-thick 
stellar wind and accretion. 

As mentioned in \S \ref{sec:intro}, the  ``immortal" 
fate of embedded stars in AGN disks is determined 
by whether the freshly-accreted hydrogen-rich disk gas can 
penetrate their
radiative layer and be fully mixed with the interior nuclear-burning 
region.  The results of our simulations show that the advective 
accretion flow is essentially halted at the outer boundary ($\sim 80 R_\odot$) of the 
convective zone (bottom panel Figure \ref{fig:snapshot}).  
Nevertheless, freshly-accreted gas is still able to 
diffuse inwards through convection-driven turbulent mixing and
sporadic inflow/outflow fluctuations,
(Figures \ref{fig:spacetime_adi} and \ref{fig:spacetime_adi_woRT})
in the convective zone.  Due to the convective overshooting across the
interface, diffusion also proceeds through
small velocity fluctuations in the lower ($\sim 25-32 R_\odot$) radiative
layer (insert in Figure \ref{fig:snapshot}).  By applying some passive-scalar 
tracers to the quasi-steady state of our fiducial case, 
we measure the effective diffusion coefficient within the lower radiative zone 
to be one per cent of that in the upper convective zone. 
Despite this difference, smoothing of composition gradient 
across the radiative zone in the envelope can still occur 
over a few thousand local dynamical timescales, which is on the order of a few years.
Note that the inner boundary of our simulation domain 
$R_{\rm in}$ is located at the top of the {\it bona fide} stellar radiative 
zone interior to $25 R_\odot$. This radiative zone extends down to the much denser 
(than the envelope) region at $\sim 5 R_\odot$ in the full MESA stellar model.  
Through this extended radiative
zone in the stellar interior, the convective overshooting effect may 
be suppressed and
the diffusion coefficient obtained from our current simulations may not be
applicable throughout the entire radiative zone. 
Moreover, the presence 
of strong outflows, particularly at higher stellar mass with $\lambda_\star 
\rightarrow 1$, most relevant to the equilibrium state of the ``immortal'' stars \citep{Jermyn2022,Chen2024}, 
can introduce additional complexities to the mixing rates. These issues need to be further investigated before assertions can be made regarding the immortality of AGN stars.

The effect of stellar rotation, not modeled in this work, may also be important in determining the accretion rate and mixing efficiency, 
potentially introducing anisotropy in the angular distribution of $L_\star$ and therefore $\lambda_{\rm diff}$  \citep{Lucy1967}, even if the background environment is relatively uniform. 
If AGN stars accrete from circumstellar disks (CSDs) just like planets from a CPD \citep{Chen2022,Li2023,Choksi2023}, 
they are expected to gain materials with large specific angular momentum and quickly spin to breakup velocity \citep{Jermyn2021}.
However, 
it's important to note that 
since radiation feedback itself provides a reduction in effective gravity in 
the fast-diffusion regime, 
the specific angular momentum of accreted materials may be scaled down 
accordingly, which can lead to modest-to-low spin growth rates for AGN stars. 
Additionally, accretion from local turbulence with chaotic spins
is expected to be prevalent in AGN disks.  Sporadic re-orientation
in the angular momentum vectors of the freshly accreted gas
may reduce any initial non-negligible stellar rotation towards  
much smaller limiting stellar spin rates \citep{ChenLin2023}. 
Analysis of these processes needs to be included in the modeling 
of AGN-disk stars' accretion and evolution, along with the 
self-consistent treatment of the above thermal-feedback effects. 

Finally, 
we remark that the general physical scalings revealed by our simulations could also have implication for the radiative feedback during metal-poor massive star formation in the early universe, as demonstrated in simulations of \citet{Hosokawa2011, Chon2024}. These stellar seeds could serve as potential precursors to supermassive stars and black holes. 
In future 1D long-term evolution modeling of these massive stars \citep[e.g.][]{Hosokawa2013}, 
rather than applying parameterized constant accretion rates,  
it may be more realistic to calculate steady-state accretion rates ``on the fly" based on stellar and environmental conditions, 
although their evolutionary paths can diverge significantly from those of 
the embedded stars in AGN disks. 


\begin{acknowledgements}
YXC wants to thank Wenrui Xu, Tianshu Wang and Zhenghao Xu for helpful discussions. We thank Mathieu Renzo for providing MESA models for our initial conditions and Philippe Yao for sharing the modified Equation of State module.
\end{acknowledgements}


\bibliography{sample631}{}
\bibliographystyle{aasjournal}



\end{CJK*}
\end{document}